\documentclass[showpacs,preprintnumbers,floatfix,nofootinbib]{revtex4}

\usepackage{dcolumn}
\usepackage{bm}
\usepackage{amsmath}
\usepackage{graphicx}
\usepackage{amsfonts}
\usepackage{amssymb}
\usepackage{placeins}

\providecommand{\U}[1]{\protect\rule{.1in}{.1in}}
\providecommand{\U}[1]{\protect\rule{.1in}{.1in}}
\input{tcilatex}

\begin{document}

\title{ Singularity Structures in Coulomb-Type Potentials in Two Body Dirac
Equations of Constraint Dynamics}
\author{Horace W. Crater\footnote{
hcrater@utsi.edu}}
\affiliation{The University of Tennessee Space Institute, Tullahoma, TN 37388 }
\author{Jin-Hee Yoon\footnote{
jinyoon@inha.ac.kr}}
\affiliation{Department of Physics, Inha University, Incheon, South Korea}
\author{Cheuk-Yin Wong\footnote{
wongc@ornl.gov}}
\affiliation{Physics Division, Oak Ridge National Laboratory, Oak Ridge, TN 37831}
\affiliation{Department of Physics, University of Tennessee, Knoxville, TN 37996}

\begin{abstract}
Two Body Dirac Equations (TBDE) of Dirac's relativistic constraint dynamics
have been successfully applied to obtain a covariant nonperturbative
description of QED and QCD bound states. Coulomb-type potentials in these
applications lead naively in other approaches to singular relativistic
corrections at short distances that require the introduction of either
perturbative treatments or smoothing parameters. We examine the
corresponding singular structures in the effective potentials of the
relativistic Schr\"odinger equation obtained from the Pauli reduction of the
TBDE. We find that the relativistic Schr\"odinger equation lead in fact to
well-behaved wave function solutions when the full potential and couplings
of the system are taken into account. The most unusual case is the coupled
triplet system with $S=1$ and $L=\{(J-1),(J+1)\}$. Without the inclusion of
the tensor coupling, the effective $S$-state potential would become
attractively singular. We show how including the tensor coupling is
essential in order that the wave functions be well-behaved at short
distances. For example, the $S$-state wave function becomes simply
proportional to the $D$-state wave function and dips sharply to zero at the
origin, unlike the usual $S$-state wave functions. Furthermore, this
behavior is similar in both QED and QCD, independent of the asymptotic
freedom behavior of the assumed QCD vector potential. Light- and heavy-quark
meson states can be described well by using a simplified
linear-plus-Coulomb-type QCD potential apportioned appropriately between
world scalar and vector potentials. We use this potential to exhibit
explicitly the origin of the large $\pi$-$\rho $ splitting and effective
chiral symmetry breaking. The TBDE formalism developed here may be used to
study quarkonia in quark-gluon plasma environments.
\end{abstract}

\pacs{ 12.39.Ki,03.65.Pm,12.39.Pn }
\maketitle

\section{Introduction}

The two-body Dirac equations we discuss in this paper are based on Dirac's
constraint formalism and a minimal interaction structure for two particles
in relative motion, first used by Todorov \cite{tod71} and confirmed by both
classical \cite{fw} and quantum field theory \cite{saz97}. The constraint
approach gives more than a sophisticated method for guessing relativistic
wave equations for systems of bound quarks or general fermion-anti-fermion
systems, since it can be readily combined with the field-theoretic machinery
of the Bethe-Salpeter equation. When used with the kernel of the
Bethe-Salpeter equation for QED, it combines weak-potential agreement in QED 
\cite{bckr} with the nonperturbative structure of the field-theoretic
eikonal approximation \cite{tod71,saz97}. The minimal interaction structure
is then automatically inherited from relativistic classical \cite{yng} and
quantum field theory \cite{saz97}.

As has been demonstrated earlier, the constraint equations correspond to a
\textquotedblleft quantum-mechanical transform\textquotedblright\ \cite
{saz85,saz92} of the Bethe-Salpeter equation (BSE). This is provided by the
two coupled Dirac equations whose fully covariant interactions are
determined by QED in the Feynman gauge \cite{cra88,bckr,saz97}. Unlike most
other truncations of the Bethe-Salpeter equation, the constraint approach
does not require the use of the awkward Coulomb gauge (whose noncovariant
nature does not allow its incorporation in covariant equations). Instead,
its expansion about the BSE naturally occurs in the covariant Feynman gauge
and is free from spurious infrared singularities that occur in the other
approaches when that gauge is used \cite{sazz}.

In QCD with flavor-independent interactions this formalism leads to
spectral results in very good agreement with most of the experimental
meson spectra\footnote{ Isoscalars such as the $\eta $, $\eta ^{\prime
}$, and $\omega $ are not included}. At the same time, as stressed in
a recent publication \cite {decay,Err08}, the formalism naturally
accounts for the perturbative results of QED bound states, when
treated in a nonperturbative manner. So far this has not been fully
replicated in any other approach. In a natural way it leads not only
to good singlet-triplet state splittings for the light as well as
heavy mesons, but also to a Goldstone behavior for the pion. By this
we mean that the numerically computed pion mass tends to zero when the
quark mass tends to zero. This is tied to the same relativistic
structures that account for the nonperturbative positronium and
muonium results \cite{decay}.

The relativistic Two-Body Dirac equations may be written as an effective
one-body wave equation \cite{tod76,cra84,cra87}. The proper formulation of
this relativistic scheme requires the successful treatment (that is a
covariant elimination) of the quantum ghost states (due to the presence of
the \textquotedblleft relative time\textquotedblright ) that first appeared
in Nakanishi's work on the Bethe-Salpeter equation \cite{nak69}. These
coupled constraint equations, known as the Two Body Dirac Equations (TBDE),
have the following important characteristics. Firstly, in the special limit
in which one of the particle masses becomes infinite, the equations reduce
to the (one-body) Dirac equation. Secondly, in the general case the Pauli
reduction of the TBDE leads to a relativistic Schr\"odinger equation, which
is the same as the non-relativistic Schr\"odinger equation in form but is
relativistic in content, including all relativistic spin and relativistic
kinematics. It displays various spin-spin, spin-orbit, tensor and Darwin
terms with energy-dependent denominators. Relativistic kinematics have been
properly taken into account to give a relation between the eigenvalue of the
relativistic Schr\"odinger equation and the invariant mass of the composite
system.

We explore several related questions in this paper. How is it that the TBDE
or its equivalent relativistic Schr\"{o}dinger equation leads to a
Goldstone-like behavior of the pion? How does this approach bypass many of
the singularities that appear in the effective potentials in other
approaches without the necessity for introducing cutoff parameters? A more
complete understanding of how the TBDE are able to accomplish this not only
will aid in a better understanding of its success in meson spectroscopy but
will assist us in its application to two-body bound states in other
environments such as those in a strongly-coupled quark gluon plasma.

In the application of TBDE to QED and QCD bound state problems, the
interaction includes Coulomb-type potentials which lead naively in
other approaches to singular relativistic corrections at short
distances (delta functions and potentials more attractive than
$-1/4r^{2}$) that require the introduction of either perturbative
treatments or smoothing parameters. In the weak potential limit in
which the potential is regarded as small compared with the masses and
center of momentum (c.m.)\ energy, the effective potentials of the
relativistic Schr\"{o}dinger equation, obtained from the Pauli
reduction of the TBDE also displays these types of singular
potentials. However, as we shall see these potentials become
nonsingular when one uses the strong potential form in which one does
not ignore the potential compared with the masses and c.m.\
energy. Nevertheless, the potential of the triplet system,
${}^{2S+1}L_{J}$, with $S=1$ and $L=(J-1)$ and $(J+1)$, remains much
more pernicious as the potential retains its singular behavior
independent of whether one regards the potential as weak or strong. We
would like to describe here these unusual singularity structures in
the potential and show how the TBDE formalism leads to wave function
solutions that are nonetheless well-behaved and physically acceptable,
when the full couplings of the system are taken into account\footnote{
We note that the finite-$r$ singularity structures that occur in the
Breit equation do not appear in the TBDE \cite{cww}.}.

In addition to our investigation of the singularity structures in TBDE, we
also wish to test a simplified QCD based potential for future applications.
Previous studies of the TBDE \cite{crater2,cra94,cra88} made use of the QCD
based Alder-Piran potential \cite{adl} giving a quite successful description
for various $q\bar{q}$ states. The Adler-Piran potential contains functions
with many terms and parameters divided into different sections of the
spatial region. On the other hand, simple potentials such as the Cornell
potential \cite{crnl} appears to be adequate for many applications, although
lacking in asymptotic freedom. It is therefore desirable to seek a
simplified potential similar to the Cornell potential, but one containing
asymptotic freedom that can be easily modified for future application of the
TBDE in other quarkonium problems. One such application is on the stability
of $q\bar{q}$ states in the quark-gluon plasma which consists of quarks and
antiquarks of different flavors, and gluons. The degree to which the
constituents of a QGP can combine to form composite entities is an important
property of the plasma.

To facilitate the application of the TBDE in terms of the equivalent
relativistic Schr\"{o}dinger equation, we present relevant useful details to
indicate how various Darwin and spin-dependent potential terms can be
constructed in Appendix C. Once the various terms of the potential have been
constructed, the solution of bound states problems in the TBDE is
mathematically just as simple as the solution for bound state problems in
non-relativistic quantum mechanics.

Accordingly, we begin in Sec.\ II with a discussion of the most often cited
QCD potentials (including a simplified model for the QCD potentials to be
used for the first time in this paper in conjunction with the TBDE for meson
spectroscopy). It is well known how a naive use of Coulomb-type potential
leads to singularity problems when one introduces relativistic spin-spin
corrections. How we avoid these problems in the TBDE is the next question
considered. In Sec.\ III, we seek out first how these problems are avoided
in the Dirac equation for an external Coulomb potential. This allows us to
explore in Sec.\ IV the parallels between the singularity structures that
occur in the Dirac and TBDE and how they each connect the different
structures of the weak-potential (or perturbative) and strong-potential (or
nonperturbative) forms of the two sets of equations. Those different
structures, although requiring different treatments, give the same spectral
results. By this we mean that a perturbative treatment of the weak-potential
forms gives the same spectral results as a nonperturbative (analytic or
numerical) treatment of the strong-potential or nonperturbative forms of the
two sets of the equations. The example we use in Sec.\ V for this parallel
discussion is the TBDE for electromagnetic interactions, QED. In Sec.\ VI we
discuss the spectral results, focusing on the pion, the $\rho$, and the
singularity structure of the TBDE for a simplified QCD potential model in
the case of $^{3}S_{1}$-${}^{3}D_{1}$ or more generally $[{{}^{3}}(J-1)_{J}]$
-$[{{}^{3}}(J+1)_{J}]$\ mixing. In Sec.\ VII we summarize the results and
discuss questions that may arise when attempts are made to apply the TBDE to
two-body bound states in a QGP.

\section{QCD Model Potentials}

Previously, the authors of \cite{crater2} used a sophisticated form of the
static quark potential developed by Adler and Piran \cite{adl}, one that has
ties at all length scales to field theoretic data. Very good agreement with
experimental quarkonium spectrum was obtained. On the other hand, in
nonrelativistic treatments the most commonly used static quark potential for
potential model studies is the Cornell potential \cite{crnl}, 
\begin{equation}
V(r)=-\frac{\alpha _{c}}{r}+br,
\end{equation}
as in \cite{Bar92,Won01}. Although not displaying asymptotic freedom, it
does give the dominant Coulomb-like behavior as well as the linear quark
confinement. Early on a model was proposed by Richardson for a static
potential which a) depends only a single scale size $\Lambda $, and b)
interpolates in a simple way between asymptotic freedom and linear
confinement \cite{rch}. Richardson's model for the static interquark
potential in momentum space is 
\begin{equation}
\tilde{V}(\mathbf{q})= -\frac{16\pi }{27}\frac{1}{\mathbf{q}^{2}\ln (1+ 
\mathbf{\ q}^{2}/\Lambda ^{2})},  \label{rcr}
\end{equation}
arising from the assumption that \ 
\begin{equation}
\tilde{V}(\mathbf{q)=}-\frac{4\alpha _{s}(\mathbf{q}^{2})}{3\mathbf{q}^{2}} 
\mathbf{,}
\end{equation}
(including the color factor $-4/3$). Asymptotic freedom requires that for $ 
\mathbf{q}^{2}/\Lambda ^{2}>>1,$ 
\begin{equation}
\alpha _{s}(\mathbf{q}^{2})\rightarrow \frac{12\pi }{27}\frac{1}{\ln ( 
\mathbf{q}^{2}/\Lambda ^{2})}.
\end{equation}
On the other hand, the property of linear confinement requires that for $
\Lambda r>>1,$ $V(r) \propto r$ or equivalently that for $\mathbf{q}
^{2}/\Lambda ^{2}<<1$ one must impose $\alpha _{s}\mathbf{(\mathbf{q}
^{2})\sim q}^{-2}$.\ The interpolation of Eq.\ (\ref{rcr}) is not tied at
all in the intermediate region and only roughly tied in the large $r$ region
to any field theoretic data. Nevertheless it provides a convenient
one-parameter form for the static quark potential. \ In coordinate space it
has the form 
\begin{equation}
V(r)=\frac{8\pi \Lambda ^{2}r}{27}-\frac{8\pi f(\Lambda r)}{27r},
\end{equation}
where $f(\Lambda r)$ is given by a complicated integral transform\footnote{
In addition to the spin independent nonrelativistic model presented in \cite
{rch} see also a relativistic extension of it given in \cite{hva}.} that
displays the asymptotic freedom behavior for $r\rightarrow 0$ of 
\begin{equation}
f(\Lambda r)\rightarrow -\frac{1}{\ln \Lambda r},
\end{equation}
while for $r\rightarrow \infty ,$ 
\begin{equation}
f(\Lambda r)\rightarrow 1.
\end{equation}
\ A simpler model, which we will apply in this paper and one which displays
the same large and small $r$ behavior is\footnote{
An earlier coordinate space form that displays asymptotic freedom as well as
linear quark confinement proposed in \cite{licht} is $V=(8\pi
/27)(1-\lambda r)^{2}/(r\ln \lambda r).$} \ 
\begin{equation}
V(r)=\frac{8\pi \Lambda ^{2}r}{27}-\frac{16\pi }{27r\ln (e^{2}+1/(\Lambda
r)^{2})}.  \label{rrich}
\end{equation}
It amounts to replacing Richardson's $f(\Lambda r)$ by $2/\ln
(e^{2}+1/(\Lambda r)^{2}),$ having the same limits. Although not giving as
good a fit to the spectra as the more closely tied QCD based potential of 
\cite{adl}, the modified form of \ Eq.\ (\ref{rrich}) which we use in this
paper (see Eq.\ (\ref{rricha}) below) does provide reasonable results for
the spectrum. Furthermore its linear-plus-Coulomb-type parametrization is
more convenient for extension of the quark model to high temperature
environments.

Problems arise in the quark model with the above potentials if their
relativistic corrections are naively grafted from semirelativistic
expressions. \ \ For example, the spin-spin interaction 
\begin{equation}
\frac{\nabla^{2}V\mathbf{\sigma}_{1}\mathbf{\cdot\sigma}_{2}}{6m_{1}m_{2}},
\label{spsp}
\end{equation}
would lead to a singular delta function potential that can only be treated
perturbatively. \ Some approaches simply include cutoff parameters so that
the Laplacian is not singular. \ How do the TBDE treat this problem?
Potential energy terms such as the above arise from the second order
reductions of those equations (the Pauli forms). Let us first examine how
such problems are treated in a very natural way in the Schr\"odinger
Pauli-form of Dirac's original wave equation. \ 

\section{Singularity Structure of Pauli-Form of the Dirac Equation}

The $q$-$\bar{q}$ interaction in Eq.\ (\ref{rrich}) contains the
color-Coulomb term that is proportional to $1/r$ and a logarithmic function
of $r$. It leads naively to singular relativistic corrections that may
render the solution singular at short distances. It is worth while to
investigate Coulomb-type potential in relativistic equations. Let us be more
precise in our definitions of singular potentials. Case \cite{case}
describes how potentials that are more attractive at the origin than $
-1/4r^{2}$ must be adjusted to maintain their self-adjoint status. Let us
call such potentials, attractive singular potentials. They include
attractive delta functions and attractive $1/r^{3}$ potentials that appear
in spin-orbit terms. Such terms must either be treated only in perturbation
theory or in cases where the coupling is strong, require adjustments, e.g.
by smoothing parameters. Calogero \cite{clgr} and Frank $et~al.$ \cite{land}
also discuss another category of potentials called repulsive singular
potentials. These are repulsive potentials that exceed an inverse quadratic
power law behavior. Strictly speaking they need not be treated using
perturbation theory, although in the case of weak potentials they are most
easily treated like that. We will discuss these more in the context below.

Let us show how the Pauli-form of the Dirac equation with a Coulomb-type
interaction contains effective potentials that are repulsively singular,
when viewed in an incomplete or perturbative context. However, when viewed
in a complete or nonperturbative context the effective potentials are
nonetheless nonsingular. We examine for simplicity the case of the Dirac
equation in a Coulomb potential ($A=-\alpha /r$) (instead of the more
complicated forms with asymptotic freedom) for stationary states, \ 
\begin{equation}
(\mathbf{\alpha \cdot p+}\beta m+A)\Psi =E\Psi .
\end{equation}
Then with 
\begin{equation}
\Psi = 
\begin{pmatrix}
\phi \\ 
\chi
\end{pmatrix}
,  \label{7}
\end{equation}
we have 
\begin{equation}
\begin{pmatrix}
m-E+A & \mathbf{\sigma \cdot p} \\ 
\mathbf{\sigma \cdot p} & -E-m+A
\end{pmatrix}
\begin{pmatrix}
\phi \\ 
\chi
\end{pmatrix}
=0.
\end{equation}
Eliminating $\chi $ we obtain 
\begin{equation}
\chi =\frac{1}{E+m-A}\mathbf{\sigma \cdot p}\phi ,  \label{9}
\end{equation}
leading to 
\begin{equation}
\mathbf{(p}^{2}-\frac{iA^{\prime }}{(E+m-A)}\mathbf{\hat{r}\cdot p+}\frac{
A^{\prime }}{r(E+m-A)}\mathbf{\sigma \cdot L)}\phi =[(E-A)^{2}-m^{2}]\phi .
\label{rp}
\end{equation}

We can eliminate the first order derivative $\mathbf{\hat{r}\cdot p~}$term
by the substitution 
\begin{equation}
\phi =F(r)\psi
\end{equation}
if one takes 
\begin{equation}
\frac{F^{\prime }}{F}=-\frac{A^{\prime }}{2(E+m-A)}.
\end{equation}
Then our equation becomes 
\begin{equation}
\mathbf{(p}^{2}+\frac{1}{2}\frac{\mathbf{\nabla }^{2}A}{(E+m-A)}+\frac{3}{4} 
\frac{\left( \mathbf{\nabla }A\right) ^{2}}{(E+m-A)^{2}}\mathbf{+}\frac{
A^{\prime }}{r(E+m-A)}\mathbf{\sigma \cdot L)}\psi =[(E-A)^{2}-m^{2}]\psi ,
\label{obde}
\end{equation}
which for a Coulomb potential becomes 
\begin{equation}
\mathbf{(p}^{2}-\frac{2E\alpha }{r}-\frac{\alpha ^{2}}{r^{2}}+\frac{2\pi
\alpha \delta ^{3}(\mathbf{r)}}{(E+m+\alpha /r)}+\frac{3}{4}\frac{\alpha
^{2} }{r^{4}(E+m+\alpha /r)^{2}}\mathbf{+}\frac{\alpha }{r^{3}(E+m+\alpha
/r) } \mathbf{\sigma \cdot L)}\psi =(E^{2}-m^{2})\psi .
\end{equation}
For $\alpha >1/2$ the inverse quadratic term would lead to an overall
attractive singular potential for $S$-states. \ If one takes the weak
potential limit in which the denominators in the three succeeding terms are
replaced by $2m$ then the spin-orbit potential is an attractive singular
potential for both coupling states and must be handled by perturbative
techniques. \ Under weak potential circumstances the delta function
potential would be treated by perturbative techniques. \ Since it is
repulsive, one could, in principle treat it in a nonperturbative way. It has
been shown, however \cite{atk}, that a nonperturbative treatment of
repulsive delta functions potentials produce no effect on bound state
energies. The repulsive $1/r^{4}$ term would require special numerical
treatments (its perturbative effects on $S$-states is ill-defined). \ Let us
now compare perturbative and nonperturbative treatments of this Pauli form.

By using the atomic units $\mathbf{r=x/}(E\alpha )$ the above equation takes
the dimensionless coordinate space form of 
\begin{align}
& \mathbf{(-}\alpha ^{2}\mathbf{\nabla }_{x}^{2}-\frac{2\alpha ^{2}}{x}- 
\frac{\alpha ^{4}}{x^{2}}+\frac{2\pi \alpha ^{4}\delta ^{3}(\mathbf{x)}}{
(1+m/E+\alpha ^{2}/x)}  \label{coul} \\
& +\frac{3}{4}\frac{\alpha ^{6}}{x^{4}(1+m/E+\alpha ^{2}/x)^{2}}\mathbf{+} 
\frac{\alpha ^{4}}{x^{3}(1+m/E+\alpha ^{2}/x)}\mathbf{\sigma \cdot L)}\psi 
\notag \\
& =[1-\left( \frac{m}{E}\right) ^{2}]\psi .  \notag
\end{align}
The standard perturbative treatment retains terms through order $\alpha ^{4}$
to arrive at the equation, 
\begin{equation}
\mathbf{(}-\alpha ^{2}\mathbf{\nabla }_{x}^{2}-\frac{2\alpha ^{2}}{x}-\frac{
\alpha ^{4}}{x^{2}}+\pi \alpha ^{4}\delta ^{3}(\mathbf{x)+}\frac{\alpha ^{4} 
}{2x^{3}}\mathbf{\sigma \cdot L)}\psi =[1-\left( \frac{m}{E}\right)
^{2}]\psi .  \label{prept}
\end{equation}
The standard semirelativistic spectral results through order $\alpha ^{4}$
can be obtained by treating this as an ordinary eigenvalue problem with the
last three terms on the left hand side as a perturbation. For the ground
state this leads to 
\begin{equation}
E=m-\frac{m\alpha ^{2}}{2}-\frac{m\alpha ^{4}}{8}+O(\alpha ^{6}).  \label{pr}
\end{equation}
Note that the $\alpha ^{6}$ term in Eq.\ (\ref{coul}) does not contribute
perturbatively to this order.

\subsection{Small $r$ Effective Potential and Wave Function Behaviors}

The $\delta ^{3}(\mathbf{r)}$ potential in Eq.\ (\ref{prept}) must only be
treated perturbatively to obtain a nonzero result. \ We know, however, that
the Dirac equation in this case can be solved analytically. \ How does that
reconcile here with the appearance of these singular potentials,
particularly the $\delta ^{3}(\mathbf{r)}$ potential? Let us restrict
ourselves here to $S$-states to make our main point. \ That would mean with $
u=\sqrt{4\pi }x\psi ~$that we must include all terms \ in 
\begin{equation}
\mathbf{(-}\alpha ^{2}\frac{d^{2}}{dx^{2}}-\frac{2\alpha ^{2}}{x}-\frac{
\alpha ^{4}}{x^{2}}+\frac{3}{4}\frac{\alpha ^{6}}{x^{2}(x(1+m/E)+\alpha
^{2})^{2}}\mathbf{)}u=[1-\left( \frac{m}{E}\right) ^{2}]u.  \label{stn}
\end{equation}
in any nonperturbative solution. \ Note that we have left out here the $
\delta ^{3}(\mathbf{x)}$ term in Eq.\ (\ref{coul}) since it, together with
the Coulomb potential in the denominator would yield a vanishing result for
its contribution (its expectation value in any well-behaved basis would give
zero). \ That means that we must have the rather unusual circumstance here
of the term which does not contribute to the weak potential form Eq.\ (\ref
{prept}) (the $\alpha ^{6}$ term) having a nonperturbative effect on the
spectrum that reproduces that of the perturbative $\delta ^{3}(\mathbf{x)}$
term (in a perturbative expansion). A set of straight-forward but tedious
manipulations show how this comes about. We first point out that the $\alpha
^{6}$ term in Eq. (\ref{coul}) has a short distance behavior of that of a
repulsive $r^{-2}$ behavior that is lower order in $\alpha $, so unlike its
weak potential $1/r^{4}$ form it is not in the category of a repulsive
singular potential. This allows a standard type of solution. One finds that
the ground state wave function and eigenvalue are given by\footnote{
At short distance we have $-\frac{d^{2}u}{dx^{2}}+(\frac{3/4-\alpha ^{2}}{
x^{2}})u=0=-\beta (\beta -1)+3/4-\alpha ^{2}$ with allowed solution $\beta =
\sqrt{1-\alpha ^{2}}+\frac{1}{2}\allowbreak \allowbreak $ . \ At long
distance ($-\frac{d^{2}u}{dx^{2}}-\frac{2}{x}-\frac{\alpha ^{2}}{x^{2}}
)u=-\lambda ^{2}u$ the allowed solution has behavior $u\sim x^{\sqrt{
1-\alpha ^{2}\text{ }}}\exp (-\lambda x)$ which forces $\gamma =-1/2.$} 
\begin{align}
u& =kx^{\beta }(x(1+m/E)+\alpha ^{2})^{\gamma }\exp (-\lambda x),  \notag \\
\beta & =\sqrt{1-\alpha ^{2}}+\frac{1}{2},  \notag \\
\gamma & =-\frac{1}{2},  \notag \\
E& =m\sqrt{1-\alpha ^{2}},  \notag \\
\lambda & =\frac{\sqrt{(m/E)^{2}-1}}{\alpha }=\frac{1}{\sqrt{1-\alpha ^{2}}}.
\end{align}
We also verify that our exact solution 
\begin{equation}
E=m\sqrt{1-\alpha ^{2}}=m\left( 1-\frac{1}{2}\alpha ^{2}-\frac{1}{8}\alpha
^{4}+O(\alpha ^{6})\right) ,
\end{equation}
agrees with the perturbative spectral results Eq.\ (\ref{pr}). Note that the
small $r$ behavior of the radial part of the wave function is 
\begin{equation}
\psi \sim x^{\sqrt{1-\alpha ^{2}}-1/2}
\end{equation}
which dips toward the origin unlike the flat behavior of the nonrelativistic
limit of the Pauli form\ or mildly singular behavior of the Dirac wave
function.

In summary, the Pauli-form (\ref{coul}) of the Dirac equation in the weak
potential approximation or the perturbative form of (\ref{prept}) includes
terms missing in the strong potential or the nonperturbative form (\ref{stn}
). Also, Eq.\ (\ref{stn}) includes terms missing in the weak potential or
the perturbative form (\ref{prept}). However, they both give rise to the
same spectral results through order $\alpha ^{4}$, with one treated in a
weak potential approximation and the other treated with no approximations
(whose spectra expansion yields the same result). \ As we shall see below,
such an unusual feature (with different parts of the equation contributing
to the perturbative and nonperturbative spectral evaluations) is also
displayed in the Pauli-form of the TBDE of constraint dynamics.

The above exercise shows that what appears as singular in a perturbative
context turns out in fact to be non-singular in a full non-perturbative
treatment. For our case of the ground state, the two approaches give the
same result up to order $\alpha ^{4}$, using different parts of the
effective interaction. Does this extend to the radially and orbitally
excited states? \ Do different parts (and approximations) of the interaction
used in reaching the spectral results lead to the same results for all high
excited states. Although we have not shown this here, it is expected to be
true since the exact (nonperturbative) Pauli form should faithfully
reproduce the exact spectral results of the first order form of the Dirac
equation. This would imply that one would expect the two approaches to give
the same results through order $\alpha ^{4}$.

\section{ The Two Body Dirac Equations of Constraint Dynamics}

Dirac constructed a quantum wave equation from a first-order wave operator
that is the matrix square-root of the corresponding Klein-Gordon operator 
\cite{di28} in order to treat a single relativistic spin-one-half particle,
free or in an external field. The TBDE of constraint dynamics extend his
construction to the system of two interacting relativistic spin-one-half
particles with quantum dynamics governed by a pair of compatible Dirac
operators acting on a single 16-component wave function. For an extensive
review of this approach, see Refs.\ \cite{cra87,bckr,cra94,crater2} and
works cited therein and \cite{unusual}. We present below a brief review.

Over thirty years ago, the relativistic constraint approach first
successfully yielded a covariant yet canonical formulation of the
relativistic two-body problem for two interacting spinless classical
particles by applying a Hamiltonian approach introduced by Dirac \cite{di64}
for handling systems with constraints. It accomplished this by introducing
two constraints thereby reducing the number of degrees of freedom of the
relativistic two-body problem to that of the corresponding nonrelativistic
problem \cite{ka75}-\cite{drz75}. By this one covariantly eliminates the
troublesome relative time and relative energy. The constraints used for this
reduction are a pair of compatible generalized mass shell constraints for
each of the two interacting spinless particles:\footnote{
We use the metric $\eta _{\mu \nu }=(-1,1,1,1)$.} $p_{i}^{2}+m_{i}^{2}+\Phi
_{i}\approx 0$.

For the case of two relativistic spin-one-half particles interacting through
four-vector and scalar potentials, the two compatible 16-component Dirac
equations (\cite{cra87,bckr,cra94,crater2}) take the form 
\begin{subequations}
\begin{align}
\mathcal{S}_{1}\psi & =\gamma _{51}(\gamma _{1}\cdot (p_{1}-\tilde{A}
_{1})+m_{1}+\tilde{S}_{1})\psi =0,  \label{tbdea} \\
\mathcal{S}_{2}\psi & =\gamma _{52}(\gamma _{2}\cdot (p_{2}-\tilde{A}
_{2})+m_{2}+\tilde{S}_{2})\psi =0,  \label{tbdeb}
\end{align}
in terms of $\mathcal{S}_{i}$ operators that in the free-particle limit
become operator square roots of the Klein-Gordon operator.

The relativistic four-vector potentials $\tilde{A}_{i}^{\mu }$ and scalar
potentials $\tilde{S}_{i}$ are effective constituent potentials that in
either limit $m_{i}\rightarrow \infty $ go over to the ordinary external
vector and scalar potentials of the light-particle's one-body Dirac
equation. The covariant spin-dependent terms in $\tilde{A}_{i}^{\mu }$ and $
\tilde{S}_{i}$ are recoil terms whose general forms are nonperturbative
consequences of the compatibility condition 
\end{subequations}
\begin{equation}
\lbrack \mathcal{S}_{1},\mathcal{S}_{2}]\psi =0.  \label{cmpt}
\end{equation}
This condition also requires that the potentials depend on the space-like
interparticle separation only through the combination 
\begin{equation}
x_{\perp }^{\mu }=(\eta ^{\mu \nu }+\hat{P}^{\mu }\hat{P}^{\nu
})(x_{1}-x_{2})_{\nu }  \label{ti}
\end{equation}
with no dependence on the relative time in the c.m.\ frame. This separation
variable is orthogonal to the total four-momentum 
\begin{equation}
P^{\mu }=p_{1}^{\mu }+p_{2}^{\mu }.
\end{equation}
$\hat{P}$ is the time-like unit vector 
\begin{equation}
\hat{P}^{\mu }\equiv P^{\mu }/w,
\end{equation}
where $w$ is the total c.m. energy (the invariant rest mass), 
\begin{equation*}
w^{2}\equiv -P^{2},
\end{equation*}
so that in the c.m. frame $\hat{P}=(1,\mathbf{0)}$ and $x_{\perp }=(0,
\mathbf{r).~}$The accompanying relative four-momentum canonically conjugate
to $x_{\perp }$ is 
\begin{equation}
\ p^{\mu }=(\epsilon _{2}p_{2}^{\mu }-\epsilon _{1}p_{2}^{\mu })/w;\mathrm{\
where}\ \epsilon _{1}+\epsilon _{2}=w,\ \epsilon _{1}-\epsilon
_{2}=(m_{1}^{2}-m_{2}^{2})/w.
\end{equation}
The $\epsilon _{i}$'s are the invariant c.m. energies of each of the
(interacting) particles. \ Another consequence of the compatibility
condition is that the relative momentum is constrained to be orthogonal to
the total four-momentum 
\begin{equation}
P\cdot p\psi =0,
\end{equation}
thus providing the conjugate covariant control on the relative energy to
that on the relative time provided by Eq. (\ref{ti}). \ One finds also that
the vector and scalar potentials are defined in terms of two invariant
functions $S(r),A(r)~$ in which $r$ is the invariant 
\begin{equation}
r\equiv \sqrt{x_{\perp }^{2}}.
\end{equation}
Those potentials have the general forms 
\begin{equation}
\tilde{A}_{i}^{\mu }=\tilde{A}_{i}^{\mu }(A(r),p_{\perp },\hat{P},w,\gamma
_{1},\gamma _{2}),~\ \tilde{S}_{i}=\tilde{S}_{i}(S(r),A(r),p_{\perp },\hat{P}
,w,\gamma _{1},\gamma _{2}).
\end{equation}

The wave operators in Eqs.\ (\ref{tbdea}) and (\ref{tbdeb}) operate on a
single 16-component spinor 
\begin{equation}
\psi=\left( 
\begin{array}{c}
\psi_{1} \\ 
\psi_{2} \\ 
\psi_{3} \\ 
\psi_{4}
\end{array}
\right)  \label{spinor}
\end{equation}
in which the $\psi_{i}$ are four-component spinors.

With compatibility ensured, this two-body formalism has many advantages over
the traditional Bethe-Salpeter equation and its numerous three dimensional
truncations. One is its simplicity. A Pauli reduction and scale
transformation (\cite{cra87,bckr,cra94,crater2}) brings these equations to
this covariant relativistic Schr\"{o}dinger equation involving a four
component spinor $\psi _{+},$ 
\begin{equation}
{\biggl (}p^{2}+\Phi _{w}(\sigma _{1},\sigma _{2},p_{\perp },A(r),S(r)){\ 
\biggr )}\psi _{+}=b^{2}(w)\psi _{+},  \label{paul1}
\end{equation}
resembling an ordinary Schr\"{o}dinger equations with the interaction term $
\Phi _{w}$ including central-potential, Darwin, spin-orbit, spin-spin, and
tensor terms. The interactions are completely local but depend explicitly on
the invariant c.m. total energy $w=-P^{2}$ . \ \ The usual invariant 
\begin{equation}
b^{2}(w)\equiv
(w^{4}-2w^{2}(m_{1}^{2}+m_{2}^{2})+(m_{1}^{2}-m_{2}^{2})^{2})/4w^{2},
\label{bb}
\end{equation}
plays the role of energy eigenvalue in this equation. This invariant is the
c.m.\ value of the square of the relative momentum expressed as a function
of the invariant mass $w$.

Note that in the limit in which one of the particles becomes very heavy,
this Schr\"{o}dinger equation turns into the one obtained by eliminating the
lower component of the ordinary one-body Dirac equation in terms of the
other component (when $S(r)=0$,$~$see Eq.\ (\ref{obde}) ). \ 

\subsection{Relativistic Schr\"odinger Equation obtained from the TBDE}

In Appendix A we outline the steps needed to obtain the explicit c.m. form
of \ Eq. (\ref{paul1}). \ That form is 
\begin{align}
& \{\mathbf{p}^{2}+2m_{w}S+S^{2}+2\varepsilon _{w}A-A^{2}+\Phi _{D}  \notag
\\
& +\mathbf{L\cdot (}\boldsymbol{\sigma}_{1}\mathbf{+}\boldsymbol{\sigma}_{2}
\mathbf{)}\Phi _{SO}+\boldsymbol{\sigma}_{1}\mathbf{\cdot \hat{r}}
\boldsymbol{\sigma}_{2}\mathbf{\cdot \hat{r}L\cdot (}\boldsymbol{\sigma}_{1}
\mathbf{+}\boldsymbol{\sigma}_{2}\mathbf{)}\Phi _{SOT}  \notag \\
& +\boldsymbol{\sigma}_{1}\mathbf{\cdot }\boldsymbol{\sigma}_{2}\Phi _{SS}+(3
\boldsymbol{\sigma}_{1}\mathbf{\cdot \hat{r}}\boldsymbol{\sigma}_{2}\mathbf{
\cdot \hat{r}-}\boldsymbol{\sigma}_{1}\mathbf{\cdot }\boldsymbol{\sigma}
_{2})\Phi _{T}  \notag \\
& +\mathbf{L\cdot (}\boldsymbol{\sigma}_{1}\mathbf{-}\boldsymbol{\sigma}_{2}
\mathbf{)\Phi }_{SOD}+i\mathbf{L\cdot }\boldsymbol{\sigma}_{1}\mathbf{\times 
}\boldsymbol{\sigma }_{2}\Phi _{SOX}\}\psi _{+}  \notag \\
& =b^{2}\psi _{+}.  \label{57}
\end{align}
The detailed \ form of the separate quasipotentials $\Phi _{i}$ are also
given in Appendix A. The subscripts of most of them are self explanatory 
\footnote{
The subscript on quasipotential $\Phi _{D}$ refers to Darwin. \ It consist
of what are called Darwin terms, those that are the two-body analogue of the
two terms to the left of the spin-orbit term in the one-body Pauli reduction
given in Eq. (\ref{obde}), and ones related by canonical transformations to
Darwin interactions \cite{fw,sch73}, momentum dependent terms arising from
retardation effects.}. \ 

After the eigenvalue $b^{2}$ of (\ref{57}) is obtained, the invariant mass
of the composite two-body system $w$ can then be obtained by inverting Eq.\
( \ref{bb}). It is given explicitly by 
\begin{equation}
w=\sqrt{b^{2}+m_{1}^{2}}+\sqrt{b^{2}+m_{2}^{2}}.
\end{equation}

In the weak potential limit in which the potential is small compared with
the masses and c.m. energy, the effective potential $\Phi $ terms of the
relativistic Schr\"{o}dinger equation (\ref{57}) contain singular
potentials. Consider for example, the case of just a vector potential. The
spin-spin term $\Phi _{SS}$ includes through the part $k(r)$ a piece $\frac{1
}{3}\nabla ^{2}\mathcal{G}$ (see Appendix A) that for weak potentials, i.e.
for $|A|<<w/2$, is equal to $\nabla ^{2}A/(3w)$. (All other parts of $\Phi
_{SS}$ are negligible for weak potentials.) When placed as a correction to
the potential in the nonrelativistic limit (so that it must be divided by $
2\mu $) its contribution would be $\Phi _{SS}/2\mu =\nabla ^{2}A\mathbf{\ }
/(6m_{1}m_{2})$ which, as discussed earlier in Eq.\ (\ref{spsp}), would
produce a singular delta function for a Coulomb-like potential. A similar
analysis would show that for weak potentials $\Phi _{D}/2\mu $ $=\nabla
^{2}A/(8\mu ^{2}),$ a repulsive singular delta function potential for the
Darwin term. For the spin-orbit term, $\Phi _{SO}/2\mu =A^{\prime
}/r(1/\left( 8\mu ^{2}\right) +1/\left( 4m_{1}m_{2}\right) )$ which when
combined with the spin-orbit eigenvalues produces either a $1/r^{3}$
attractive or repulsive singular potentials. One finds a similar behavior
for the spin-orbit difference term. For the tensor terms one would find the
combination $(\Phi _{SOT}-\Phi _{T})/2\mu $ $=(A^{\prime \prime }-\frac{
2A^{\prime }}{r})/(12m_{1}m_{2})~$appearing in all angular momentum states
producing attractive singular $1/r^{3}$ potentials. (See radial equations (
\ref{pp})-(\ref{mi}) below.) As we shall see below, when the strong
potential structures are included, those type of singular potential
structures disappear. However for the tensor terms we shall find below a
remnant attractive singular potential ($\sim r^{-2.5}$) whose effect,
however, is compensated by the tensor coupling.

In the above equation (\ref{57}) one has the strong potential form if one
does not ignore the potential compared with the masses or the c.m. energy in
the expression for the $\Phi $'s. \ In the following section we demonstrate
that these strong potential forms and the unusual singularity structures in
them lead to wave function solutions that are well-behaved and physically
acceptable when the full couplings of the system are taken into account. \
We do this by examining the short distance behavior of these equations.

\section{The Behavior of Singlet and Triplet Wave Functions at Small $r$ in
QED}

In QED bound state problems, as has been shown in detail in \cite{bckr} for $
A=-\alpha /r$ the nonperturbative (numerical) treatment of the TBDE give the
same spectral results as a perturbative treatment of the weak potential form
(which in turn gave the same results as older standard methods). We note
that this agreement was for a number of radial and orbitally excited states
for equal and unequal mass as well as the ground state. In this section we
examine analytically the structures of the four-component form of Eq.\ (\ref
{57}).

For the equal mass singlet states, there is no spin-mixing and with $
A=-\alpha /r,~S=0$ we obtain 
\begin{equation}
\left [-\frac{d^{2}}{dr^{2}}-\frac{2\varepsilon _{w}\alpha }{r}-\frac{\alpha
^{2} }{r^{2}}\right ]v_{0}=b^{2}v_{0}.
\end{equation}
As was shown in \cite{van} where this equation first appeared, it has an
exact solution with the eigenvalue given by 
\begin{align}
w& =m\sqrt{2+2 \bigg /\sqrt{1+{\alpha ^{2}} \bigg / \biggl ( n+\sqrt{(L+ 
\frac{1}{2})^{2}-\alpha ^{2}}-L-\frac{1}{2} \biggr )^{2}}}  \notag \\
& =2m-m{\alpha ^{2}}/{4}n^{2}-(m\alpha ^{4}/2n^{3})[1/(2L+1)-11/32n]+O(\alpha
^{6}),  \label{exct}
\end{align}
or $w=2m-m{\alpha ^{2}}/{4}-21m\alpha ^{4}/64$ for the ground state. \ The
small $r$ wave radial function behaviors $v_{0}\sim r^{\frac{1}{2}\sqrt{
1-4\alpha ^{2}}+\frac{1}{2}}$ which implies the mildly singular but
physically acceptable behavior of $\psi _{0}=v_{0}/r\sim r^{\frac{1}{2}\sqrt{
1-4\alpha ^{2}}-\frac{1}{2}}$. Of course, its perturbative treatment gives
the same results as the exact one through order $\alpha ^{4}$.

The ground state triplet c.m.\ energy (excluding the annihilation
contribution) perturbatively is \cite{bckr} $w=2m-m{\alpha ^{2}}/{4}-m\alpha
^{4}/192.$ This result, unlike the singlet case has not been obtained as an
expansion of an exact analytic result. Unlike the one-body Dirac equation,
the triplet states do not possess a known exact spectral solution for the
TBDE. However, it has been verified that a nonperturbative (numerical)
solution of the TBDE does produce a result that agrees with the perturbative
evaluation \cite{bckr}.

We are more interested here in the behavior of the potentials in Eqs.\ (\ref
{pl}) and (\ref{mi}) and the resultant wave functions at small $r$. Those,
we show, can be determined analytically and provide a severe test for the
strong potential forms of the effective potentials. By using the effective
potentials in those equations we find that at small $r$ \ Eqs.\ (\ref{pl})
and (\ref{mi}) become ($A=-\alpha /r,~S=0$) for a general $J$ 
\begin{align}
& \left[ -\frac{d^{2}}{dr^{2}}-\frac{\alpha ^{2}}{r^{2}}+\frac{(J+1)(J-\frac{
1}{4(2J+1)})}{r^{2}}-\frac{2J(J+1)}{2J+1}\frac{1}{r^{2}}\sqrt{\frac{\alpha }{
2rw}}\right] u_{+}  \notag  \label{sing1} \\
& +\frac{2\sqrt{J(J+1)}}{2J+1}\left[ -\frac{1}{8r^{2}}+\frac{J+1}{r^{2}}
\sqrt{\frac{\alpha }{2rw}}\right] u_{-}=0,
\end{align}
and 
\begin{align}
& \left[ -\frac{d^{2}}{dr^{2}}-\frac{\alpha ^{2}}{r^{2}}+\frac{J(J+1-\frac{1
}{4(2J+1)})}{r^{2}}+\frac{2J(J+1)}{(2J+1)r^{2}}\sqrt{\frac{\alpha }{2rw}}
\right] u_{-}  \notag  \label{sing2} \\
& +\frac{2\sqrt{J(J+1)}}{2J+1}\left[ -\frac{1}{8r^{2}}-\frac{J}{r^{2}}\sqrt{
\frac{\alpha }{2rw}}\right] u_{+}=0.
\end{align}
Note that in Eq.\ (\ref{sing1}), the potential for the $u_{+}$ wave function
contains an attractive term that is proportional to $r^{-5/2}$. Without the
coupling to the $u_{-}$ wave function, Eq.\ (\ref{sing1}) would lead to an
attractive singular potential. However, the potential for the $u_{-}$ wave
function in Eq.\ (\ref{sing2}) contains a repulsive term that is
proportional to $r^{-5/2}$. The wave function $u_{+}$ is thus prevented from
collapsing to the center due to the coupling of $u_{+}$ to $u_{-}$ in Eq.\ (
\ref{sing1}). In fact, because of the coupling, the short distance behavior
of $u_{+}$ and $u_{-}$ become proportional and given by 
\begin{align}
u_{+}(r)& =r^{\lambda },  \notag \\
u_{-}(r)& =\frac{Jr^{\lambda }}{\sqrt{J(J+1)}}=\frac{Ju_{+}(r)}{\sqrt{J(J+1)}
},  \label{rlt}
\end{align}
which lead to an exact cancellation of the singular attractive $r^{-5/2}$
terms in Eqs.\ (\ref{sing1}) and (\ref{sing2}) with their repulsive $
r^{-5/2} $ counterparts. The power index $\lambda $ with the correct
physical behavior at the origin is 
\begin{equation}
\lambda =\sqrt{J(J+1)-\alpha ^{2}}+\frac{1}{2},
\end{equation}
so that at short distance the correct physical behavior of the wave
functions is 
\begin{align}
u_{+}(r)& =r^{(1/2+\sqrt{J(J+1)-\alpha ^{2}})},  \notag \\
u_{-}(r)& =\frac{Jr^{(1/2+\sqrt{J(J+1)-\alpha ^{2}})}}{\sqrt{J(J+1)}}.
\end{align}
The corresponding radial parts of the wave functions would be 
\begin{align}
\psi _{+}(r)& =r^{\sqrt{J(J+1)-\alpha ^{2}}-\frac{1}{2}},  \notag \\
\psi _{-}(r)& =\frac{Jr^{\sqrt{J(J+1)-\alpha ^{2}}-\frac{1}{2}}}{\sqrt{J(J+1)
}}.  \label{dip}
\end{align}
Focusing on $J=1$, both of these wave functions would show an unusual
feature of a dip not only in the $D$ state but also in the $S$ state. This
is in contrast to what occurs in the singlet case where there is the mildly
singular behavior or in the nonrelativistic case where the behavior is flat.
Also, whereas an upper bound of $\alpha =1/2$ is placed on the coupling in
the singlet case, here the upper bound is $\sqrt{2}$ for well-behaved wave
functions. The unusual dips of the $S$- and $D$-wave functions for QED also
shows up in numerical solution of the QCD case, as will be seen in Figs.\ 3
and 4.

For the special case of $J=0$ ($^{3}P_{0}$) the short distance behavior is 
\begin{equation}
\{-\frac{d^{2}}{dr^{2}}-\frac{\alpha^{2}}{r^{2}}\}u_{0}=0,
\end{equation}
just as it is for the $^{1}S_{0}$ case.

For the coupled $^{1}J_{J},^{3}J_{J}$ states, there will be spin mixing in
the general case of unequal constituent masses. For equal mass or $J=0$ the
equations decouple. However, in the limit of small $r$, the mixing term
vanishes anyway, leaving us with the uncoupled short distance behavior for
the wave equations (\ref{ss}) and (\ref{pp}) for $J>0~$ given by 
\begin{align}
\{-\frac{d^{2}}{dr^{2}}+\frac{J(J+1)}{r^{2}}-\frac{\alpha ^{2}}{r^{2}}
\}v_{S=0}& =0,  \notag \\
\{-\frac{d^{2}}{dr^{2}}+\frac{J(J+1)}{r^{2}}-\frac{\alpha ^{2}}{r^{2}}-\frac{
1}{4r^{2}}\}v_{S=1}& =0.
\end{align}
The radial wave functions behave as 
\begin{align}
v_{S=0}& \rightarrow r^{(1/2+\sqrt{(J+1/2)^{2}-\alpha ^{2}})},  \notag \\
v_{S=1}& \rightarrow r^{(1/2+\sqrt{J(J+1)-\alpha ^{2}})}.
\end{align}
The corresponding radial parts of the total wave functions would be 
\begin{align}
\psi _{S=0}& \rightarrow r^{(-1/2+\sqrt{(J+1/2)^{2}-\alpha ^{2}})},  \notag
\\
\psi _{S=1}& \rightarrow r^{(-1/2+\sqrt{J(J+1)-\alpha ^{2}})},
\end{align}
and have acceptable behavior for $J>0$ when $\alpha $ is bounded by $J+1/2$
and $\sqrt{J(J+1)}$ respectively. Notice that unlike the tensor mixing case,
there is no short distance connection in the spin mixing case between the
wave function scales as in Eq.\ (\ref{rlt}). For $J=0$, only the first wave
function is relevant and the wave function is well-behaved at short
distances for $\alpha $ bounded by $1/2$.

We can summarize the short-distance behavior of the wave function for
different spin and angular momentum states. For the coupled or uncoupled $
({}^{1}J_{J},{}^{3}J_{J})$ states, the corresponding potentials are not
singular. In the tensor coupling case, there are both attractive and
repulsive singular potentials. However, the effect of the coupled equations
is to produce eventually well-behaved wave functions. Thus all wave
functions at short-distance are well-behaved for all spin and angular
momentum states for appropriate $\alpha $ bounds.

Next we examine the singularity structure of the TBDE in the case of QCD
bound states. \ The short distance behaviors (which we attribute to the
invariant $A(r)$) of QED and QCD are well known to have a crucial
distinction. \ Renormalization group arguments show that in QED the
asymptotic behavior displays a singularity structure\footnote{
The running coupling constant in spinor QED is given by \cite{quigg} $
\alpha_{R}(q^{2})\rightarrow\alpha
_{R}(m^{2})/(1-\alpha_{R}(m^{2})\ln(-q^{2}/m^{2})/3\pi).$} at finite (though
very large) energy that invalidates perturbation theory \cite{wein}. In
contrast similar renormalization group arguments show for QCD the asymptotic
behavior displays a structure\footnote{
The running coupling constant in QCD is given by \cite{quigg} $
1/\alpha_{s}(q^{2})\rightarrow1/\alpha_{s}(\mu
^{2})+(33-2n_{f})\ln(-q^{2}/\mu^{2})/12\pi.$} that at high energy strongly
validates perturbation theory (asymptotic freedom) \cite{wein}. For this
reason we have not included a running coupling constant above in the QED
application.

\section{Simple QCD Based Model for Quark-Anti-Quark Bound States.}

In \cite{crater2} a fully relativistic calculation of the meson spectrum was
made using the above TBDE with the invariants $A(r)$ and $S(r)$ determined
from a relativistic extension of the nonrelativistic Adler-Piran \cite{adl}
static quark potential. \ That paper investigated how well the relativistic
constraint approach performs in comparison with selected alternatives
including those of Godfrey and Isgur \cite{isgr} when used to produce a
single fit of experimental results over the whole meson spectrum. The
authors of \cite{crater2} found that the fit provided by the two-body Dirac
model for the entire meson spectrum competes with the best fits to partial
spectra provided by the others and does so with a smaller number of
interaction functions. Furthermore this is done without additional cutoff
parameters necessary to make some of the other approaches numerically
tractable.

In this section we examine the spectral results produced by a relativistic
version of a simpler model motivated by the static quark potential of
Richardson \cite{rch}. It is not our purpose here to improve upon the
results of \cite{crater2} but rather to take advantage of the simpler
structure of the model to explore certain details of the effective potential
and wave function behaviors generated by the TBDE along the lines discussed
in the above sections for QED. This will also allow us to more readily make
changes in the invariants $A(r)$ and $S(r)$ determined from this simpler
model to facilitate investigations on quarkonium stability in the presence
of a quark-gluon plasma.

\subsection{The Model and Spectral Results}

For our potential model we consider here a slightly more generalized form of
Eq.\ (\ref{rrich}), 
\begin{equation}
V(r)=\frac{8\pi \Lambda ^{2}r}{27}-\frac{16\pi }{27r\ln (Ke^{2}+B/(\Lambda
r)^{2})}.  \label{rricha}
\end{equation}
This has the feature of giving a variable parameter $K$ to the long distance
Coulomb behavior as well as (through $B$) an effective QCD parameter $
\Lambda $ distinct from the one placed in the linear potential. \ In our QCD
spectral work we assign the confining piece of this potential to the
invariant $S~$that controls the scalar potential and the Coulomb-like piece
to the invariant $A$ that controls the vector potential (see Eqs.\ (\ref{kg}
),(\ref{scl}),(\ref{vectl}), and (\ref{vec})). \ Thus with $r=\sqrt{x_{\perp
}^{2}}$ the equations 
\begin{align}
S(r)& =\frac{8\pi \Lambda ^{2}r}{27},  \notag \\
A(r)& =-\frac{16\pi }{27r\ln (Ke^{2}+B/(\Lambda r)^{2})} +\frac{e_{1}e_{2}}{
4\pi r} ,  \label{sa}
\end{align}
together with the relativistic Schr\"odinger equation (\ref{57}) of our TBDE
(\ref{b2}) define the covariant formalism for our QCD spectral work ($
e_{1},e_{2}$ are the respective electric charges of the quark and
anti-quark). \ With the parameter values listed in Table I we obtain the
following spectral results shown in Table II and Figure 1. The mass of the $
u $ and $d$ quarks are only 55 MeV. We observe that the masses of the
low-lying mesons are well reproduced in the TBDE treatment with a minimum
number of parameters. The singlet-triplet splittings of $\pi $-$\rho $ and $
\eta _{c}$-$J/\psi $ are well reproduced. The agreement of the experimental
masses with theory is not as impressive as that using the Adler-Piran
potential. (Note that as with the Adler-Piran potential, the ground state
singlet/triplet splitting for the charmonium system appears too large, while
that between the ground and first excited states appears too small. The same
problem also occurs for the bottomonium system with the recently observed $
\eta _{b}$ \cite{aub}.) The simplified potential has the advantage of
simplicity and ease of adoptive modification that can be useful for
applications of the TBDE to other quarkonium problems.

\begin{figure}[h]
\includegraphics[angle=0,scale=0.50]{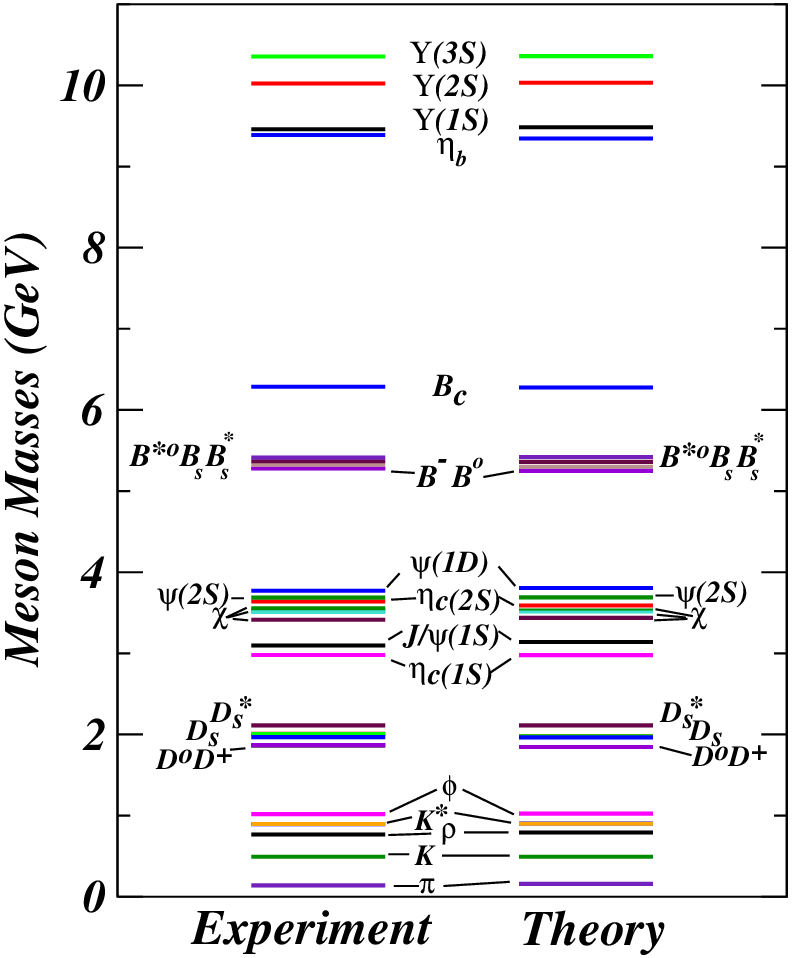} \vspace*{0.0cm} 
\caption{ (Color online) Comparison of experimental and theoretical masses
obtained with the Two-Body Dirac Equations.}
\end{figure}

We next consider some interesting and unexpected behaviors of the solutions
of the equations in the next few subsections.

\begin{table}[h]
\centering $ 
\begin{tabular}{|l|}
\hline
$\Lambda=0.4218~\text{GeV}$ \\ \hline
$B=0.05081$ \\ \hline
$K=4.198$ \\ \hline
$m_{u}=0.0557~\text{GeV}$ \\ \hline
$m_{d}=0.0553~\text{GeV}$ \\ \hline
$m_{s}=0.2499~\text{GeV}$ \\ \hline
$m_{c}=1.476~\text{GeV}$ \\ \hline
$m_{b}=4.844~\text{GeV}$ \\ \hline
\end{tabular}
\ \ $
\caption{Quark Model Parameter Values }
\label{key}
\end{table}

\begin{table}[h]
\centering                                 
\begin{tabular}{|lllll|}
\hline
Meson~~~~~~~~~~~~~~~~~~~ & Exp.(GeV) & Theory(GeV) & Exp.-Theory(GeV) &  \\ 
\hline
$\pi:u\overline{d}\ 1\, {}^{1}S_{0}$ & 0.140 & 0.159 & -0.019 &  \\ 
$\rho:u\overline{d}\ 1\, {}^{3}S_{1}$ & 0.775 & 0.792 & -0.017 &  \\ \hline
$K\, {}^{-}:s\overline{u}\ 1\, {}^{1}S_{0}$ & 0.494 & 0.493 & ~0.001 &  \\ 
$K^{0}:s\overline{d}\ 1\, {}^{1}S_{0}$ & 0.498 & 0.488 & ~0.010 &  \\ 
$K^{\ast}:s\overline{u}\ 1\, {}^{3}S_{1}$ & 0.892 & 0.903 & -0.011 &  \\ 
$K^{\ast}:s\overline{d}\ 1\, {}^{3}S_{1}$ & 0.896 & 0.901 & -0.005 &  \\ 
\hline
$\phi:s\overline{s}\ 1\, {}^{3}S_{1}${\LARGE \ } & 1.019 & 1.025 & -0.006 & 
\\ \hline
$D^{0}:c\overline{u}\ 1\, {}^{1}S_{0}$ & 1.865 & 1.840 & ~0.025 &  \\ 
$D^{+}:c\overline{d}\ 1\, {}^{1}S_{0}$ & 1.870 & 1.845 & ~0.025 &  \\ 
$D^{\ast0}:c\overline{u}\ 1\, {}^{3}S_{1}$ & 2.010 & 1.981 & ~0.029 &  \\ 
$D^{\ast+}:c\overline{d}\ 1\, {}^{3}S_{1}$ & 2.007 & 1.979 & ~0.028 &  \\ 
$D_{s}:c\overline{s}\ 1\, {}^{1}S_{0}$ & 1.968 & 1.965 & ~0.003 &  \\ 
$D_{s}^{\ast}:c\overline{s}\ 1\, {}^{3}S_{1}$ & 2.112 & 2.112 & ~0.000 &  \\ 
\hline
$\eta_{c}:c\overline{c}\ 1\, {}^{1}S_{0}$ & 2.980 & 2.978 & ~0.002 &  \\ 
$J/\psi(1S):c\overline{c}\ 1\, {}^{3}S_{1}$ & 3.097 & 3.140 & -0.043 &  \\ 
$\psi(2S):c\overline{c}\ 2\, {}^{3}S_{1}$ & 3.686 & 3.689 & -0.003 &  \\ 
$h_{1}:c\overline{c}\ 1\, {}^{1}P_{1}$ & 3.526 & 3.522 & ~0.004 &  \\ 
$\chi_{0}:c\overline{c}\ 1\, {}^{3}P_{0}$ & 3.415 & 3.436 & -0.021 &  \\ 
$\chi_{1}:c\overline{c}\ 1\, {}^{3}P_{1}$ & 3.511 & 3.515 & -0.004 &  \\ 
$\chi_{2}:c\overline{c}\ 1\, {}^{3}P_{2}$ & 3.556 & 3.541 & ~0.015 &  \\ 
$\eta_{c}:c\overline{c}\ 2\, {}^{1}S_{0}$ & 3.638 & 3.591 & ~0.047 &  \\ 
$\psi(1D):c\overline{c}\ 1\, {}^{3}D_{1}$ & 3.773 & 3.804 & -0.031 &  \\ 
\hline
$B^{-}:b\overline{u}\ 1\, {}^{1}S_{0}$ & 5.279 & 5.249 & ~0.030 &  \\ 
$B^{0}:b\overline{d}\ 1\, {}^{1}S_{0}$ & 5.280 & 5.248 & ~0.032 &  \\ 
$B^{\ast0}:b\overline{u}\ 1\, {}^{3}S_{1}$ & 5.325 & 5.299 & ~0.026 &  \\ 
$B_{s}:b\overline{s}\ 1\, {}^{1}S_{0}$ & 5.366 & 5.360 & ~0.006 &  \\ 
$B_{s}^{\ast}:b\overline{s}\ 1\, {}^{3}S_{1}$ & 5.413 & 5.420 & -0.007 &  \\ 
$B_{c}^{-}~b\overline{c}\ 1\, {}^{1}S_{0}$ & 6.276 & 6.276 & ~0.000 &  \\ 
\hline
$\eta_{b}:b\overline{b}\ 1\, {}^{1}S_{0}$~~~~~ & 9.389 & 9.345 & ~0.044 & 
\\ 
$\Upsilon(1S):b\overline{b}\ 1\, {}^{3}S_{1}$ & 9.460 & 9.484 & -0.024 &  \\ 
$\Upsilon(2S):b\overline{b}\ 2\, {}^{3}S_{1}$ & 10.023 & 10.033 & -0.010 & 
\\ 
$\Upsilon(3S):b\overline{b}\ 3\, {}^{3}S_{1}$ & 10.355 & 10.360 & -0.005 & 
\\ \hline
\end{tabular}
\caption{Selected Portions of Meson Spectrum }
\end{table}

\subsection{Behaviors of singlet and triplet solution to bound state
equations}

\subsubsection{Detailed Analysis of the Pion and Rho Bound States.}

One of the most unusual features of the spectral results of the TBDE is the
fairly accurate production of the large $\pi $-$\rho $ mass splitting. \
First we point out that unlike most other potential models, the up and down
quark masses do not take on the values typically seen of about $300$ MeV in
most other potential models (see e.g. \cite{Won01}). \ Just as with the more
detailed Adler-Piran potential \cite{adl}, the quark masses that give the
best fit in our model here are on the order of $50-60$ MeV. \ It is of
interest to see how the small pion mass comes about. In earlier work \cite
{crater2}, it was shown numerically that as the quark masses tend to zero,
so also does the bound state mass. Thus our analysis here bears on the
dynamics of chiral symmetry breaking. The bound state equation for $S$
states has the general form of Eq.\ (\ref{ss}) and in this case reduces to 
\begin{align}
& \{-\frac{d^{2}}{dr^{2}}+2m_{w}S+S^{2}+2\varepsilon _{w}A-A^{2}+\Phi _{D}
\mathbf{-}3\Phi _{SS}\}v_{0}  \notag  \label{eq71} \\
& \equiv \{-\frac{d^{2}}{dr^{2}}+\Phi \}v_{0}\equiv \{-\frac{d^{2}}{dr^{2}}
+2m_{w}S+S^{2}+2\varepsilon _{w}A-A^{2}+\Phi _{SD}\}  \notag \\
& \equiv \{-\frac{d^{2}}{dr^{2}}+\Phi _{SI}+\Phi _{SD}\}=b^{2}v_{0}.
\end{align}
Table III below lists the expectation values of the various parts
(units in GeV$^{2}$).  Even though the quark masses make up a
substantial portion of the pion mass it would  be a gross
mis-statement to say that the pion\ bound state is nearly
nonrelativistic. The constituent kinetic portion $ \langle
-\frac{d^{2}}{dr^{2}}\rangle $ and potential portion $\langle \Phi
\rangle $ are huge compared with the rest mass squared (0.00302
GeV$^{2}$) but nearly cancel, leaving a small $b^{2}$ which
corresponds to a pion mass of $0.159~$ GeV. The pion as a
quark-antiquark system has large kinetic energies and potential
energies that counterbalance each other. When one looks at the
potential energy $\langle \Phi \rangle =-0.8475$ GeV$^{2}$ from
various contributions, one notes that the spin-independent
contribution $\langle \Phi _{SI}\rangle =\langle 2m_{w}S\rangle
+\langle S^{2}\rangle +\langle 2\epsilon _{w}A\rangle -\langle
A^{2}\rangle =-0.3832$ GeV$^{2}$ while the combined Darwin and
spin-dependent contributions are $\langle \Phi _{SD}\rangle =\langle
\Phi _{D}-3\Phi _{SS}\rangle =$-0.4643 GeV$^{2}$. They are of the same
order of magnitude. Among the two huge contributions to $ \langle \Phi
_{SD}\rangle $, the Darwin term $\langle \Phi _{D}\rangle $ and
$\langle -3\Phi _{SS}\rangle $ nearly cancel one another. The small
mass of the pion arises in no small measure from delicate
cancellations of these very large contributions from the Darwin and
spin-spin interactions, as well as the balance of kinetic energy and
potential energies. Such does not occur with the other mesons. (See
for example the rho meson below). As seen in Fig. 2 there is nothing
unusual in the pion wave function in constraint dynamics.

\begin{figure}[h]
\includegraphics[angle=0,scale=0.50]{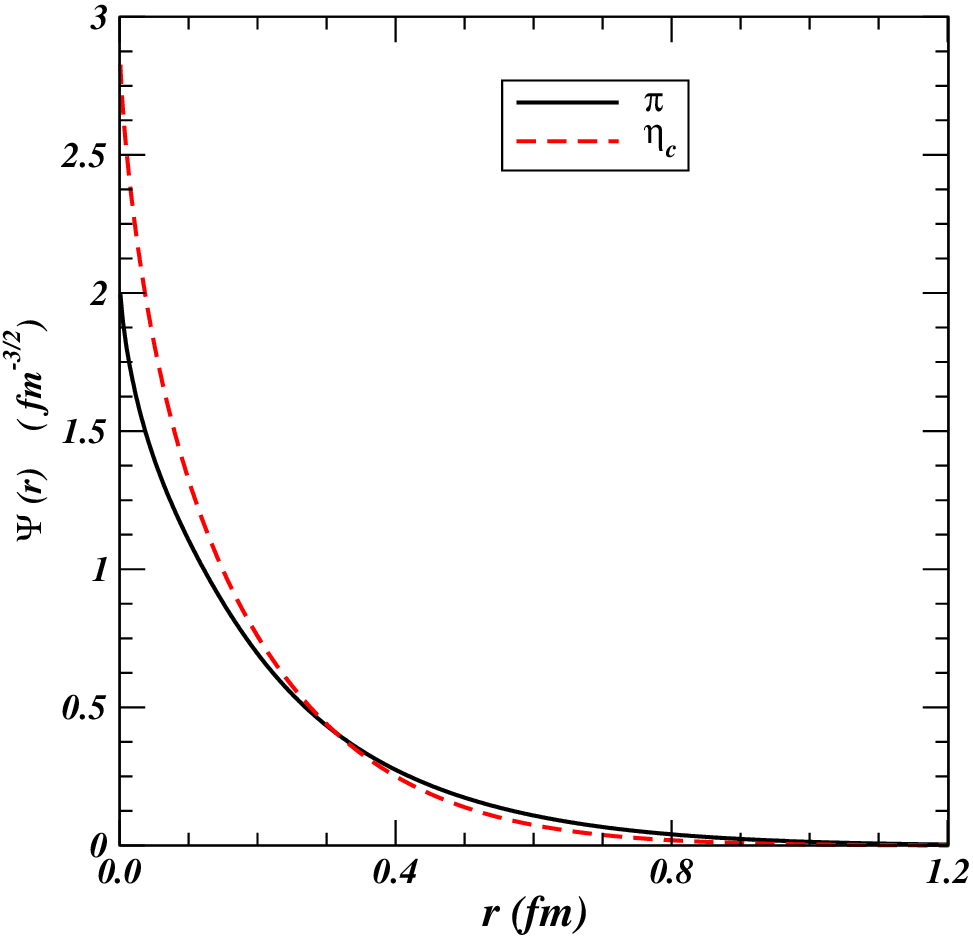} 
\caption{(Color online) Wave functions of $\protect\pi $ and $\protect\eta 
_{c}$ mesons}
\end{figure}
\begin{table}[h]
\begin{tabular}{|l|l|}
\hline
$\langle -\frac{d^{2}}{dr^{2}}\rangle $ & 0.8508 \\ 
$\langle \Phi \rangle $ & -0.8475 \\ \hline
$\langle 2m_{w}S\rangle $ & 0.0103 \\ \hline\hline
$\langle S^{2}\rangle $ & 0.0942 \\ \hline
$\langle 2\varepsilon _{w}A\rangle $ & -0.0598 \\ \hline
$\langle -A^{2}\rangle $ & -0.4279 \\ \hline
$\langle \Phi _{SI}\rangle $ & -0.3832 \\ \hline\hline
$\langle \Phi _{D}\rangle $ & -3.804 \\ \hline
$\langle \mathbf{-}3\Phi _{SS}\rangle $ & 3.340 \\ \hline
$\langle \Phi _{SD}\rangle $ & -0.4643 \\ \hline\hline
$\langle b^{2}\rangle $ & 0.0033 \\ \hline
\end{tabular}
\caption{Expectation values of various terms in (\protect\ref{eq71}) (in GeV$
^{2}$) that contribute to the pion eigenvalue equation (\protect\ref{eq71}).}
\end{table}

By contrast now we present a similar table for the $\rho $ meson. \ The
coupled equations are

\begin{align}
& \{-\frac{d^{2}}{dr^{2}}+2m_{w}S+S^{2}+2\varepsilon _{w}A-A^{2}+\Phi
_{D}+\Phi _{SS}\}u_{+}  \notag \\
& +\frac{2\sqrt{2}}{3}\{3\Phi _{T}-6\Phi _{SOT}\}u_{-}  \notag \\
& \equiv \{-\frac{d^{2}}{dr^{2}}+\Phi _{++}\}u_{+}+\Phi _{+-}u_{-}  \notag \\
& =b^{2}u_{+},  \label{rhos}
\end{align}
$\allowbreak \allowbreak \allowbreak $ and

\begin{align}
& \{-\frac{d^{2}}{dr^{2}}+\frac{6}{r^{2}}+2m_{w}S+S^{2}+2\varepsilon
_{w}A-A^{2}+\Phi _{D}-6\Phi _{SO}+\Phi _{SS}-2\Phi _{T}+2\Phi _{SOT}\}u_{-} 
\notag \\
& +\frac{2\sqrt{2}}{3}\{3\Phi _{T}\}u_{+}  \notag \\
& \equiv \{-\frac{d^{2}}{dr^{2}}+\frac{6}{r^{2}}+\Phi _{--}\}u_{-}+\Phi
_{-+}u_{+}  \notag \\
& =b^{2}u_{-}.  \label{rhod}
\end{align}
In terms of expectation values we have 
\begin{eqnarray}
b^{2} &=&\int_{0}^{\infty }dr[u_{+}(r)\left( -\frac{d^{2}}{dr^{2}}+\Phi
_{++}\right) u_{+}(r)+u_{+}(r)\Phi _{+-}u_{-}(r)  \notag \\
&&+u_{-}(r)\left( -\frac{d^{2}}{dr^{2}}+\frac{6}{r^{2}}+\Phi _{--}\right)
u_{-}(r)+u_{-}(r)\Phi _{-+}u_{+}(r)]  \label{rho}
\end{eqnarray}
Table IV below lists the expectation values of the various parts (units in
GeV$^{2}$). \ \ The quark masses make up only a small portion of the rho
mass. As with the pion, the rho is highly relativistic, but unlike the pion,
its relativistic nature is not hidden in large cancellations.\ The
constituent kinetic portion $\langle -\frac{d^{2}}{dr^{2}}\rangle
_{++},\langle -\frac{d^{2}}{dr^{2}}+\frac{6}{r^{2}}\rangle _{--},$ and
potential portions $\langle \Phi \rangle _{++},\langle \Phi \rangle
_{+-}\langle \Phi \rangle _{-+},\langle \Phi \rangle _{--}$ are huge
compared with the rest mass squared (0.00302 GeV$^{2}$). \ There is some
cancellation but not nearly to the extent that occurs in the pion. There is
left a significant $b^{2}$ $=$0.1411 GeV$^{2}~$which corresponds to a
computed rho mass of 0.796 GeV. Note that by itself, the $S$-wave portion
(kinetic plus potential) is negative and large compared to the rest mass. \
Thus the positive $D$-wave portion is crucial to bring the rho mass in line
with the observed value. The two spin-independent contributions are $\langle
\Phi_{SI}\rangle _{ii} =\langle 2m_{w}S\rangle _{ii}+\langle S^{2}\rangle
_{ii}+\langle 2\epsilon_{w}A\rangle _{ii}-\langle A^{2}\rangle _{ii}$ where $
{ii}=++$ and $--$. The magnitudes of $\langle \Phi_{SI}\rangle _{++}=-0.1680$
GeV$^2$ and $\langle \Phi_{SI}\rangle _{--}=0.01461~$ GeV$^{2}$ differ by an
order of magnitude, as do the diagonal Darwin and spin-dependent
contributions $\langle \Phi _{SD}\rangle _{++}=\langle \Phi _{D}+\Phi
_{SS}\rangle _{++}=$-0.3154 GeV$^{2}$ and $\langle \Phi _{SD}\rangle
_{--}=\langle \Phi _{D}+\Phi _{SS}-2\Phi _{T}-6\Phi _{S0}+2\Phi
_{S0T}\rangle _{--}=$ 0.03461 GeV$^{2}$. The two sets differ by roughly a
factor of 2 and are of the same sign. The off diagonal tensor terms $\langle
\Phi \rangle _{+-}=\frac{2\sqrt{2}}{3}\langle 3\Phi _{T}-6\Phi _{SOT}\rangle
_{+-},\langle \Phi \rangle _{-+}=\frac{2\sqrt{2}}{3} \langle 3\Phi
_{T}\rangle _{-+}$ are both quite large but of opposite sign so their
overall effects almost cancel. \ 

\begin{table}[h]
\centering              
\begin{tabular}{|l|ll|}
\hline
~~~$\langle -\frac{d^{2}}{dr^{2}}\rangle _{++},\langle -\frac{d^{2}}{dr^{2}}
+ \frac{6}{r^{2}}\rangle _{--}$ & 0.3085 & 0.2812 \\ \hline
$\langle \Phi \rangle _{++},~\langle \Phi \rangle _{--}$ & -0.4835 & 0.04923
\\ \hline
$\langle \Phi \rangle _{+-},~\langle \Phi \rangle _{-+}$ & 0.3090 & -0.3088
\\ \hline
$\langle 2m_{w}S\rangle _{++},~\langle 2m_{w}S\rangle _{--}$ & 0.00263 & 
0.000571 \\ \hline\hline
$\langle S^{2}\rangle _{++},~\langle S^{2}\rangle _{--}$ & 0.1631 & 0.04457
\\ \hline
$\langle 2\varepsilon _{w}A\rangle _{++},~\langle 2\varepsilon _{w}A\rangle
_{--}$ & -0.2091 & -0.02109 \\ \hline
$\langle -A^{2}\rangle _{++},~\langle -A^{2}\rangle _{--}$ & -0.1247 & 
--.00944 \\ \hline
$\langle \Phi _{SI}\rangle _{++},~\langle \Phi _{SI}\rangle _{--}$ & -0.1680
& 0.01461 \\ \hline\hline
$\langle \Phi _{D}\rangle _{++},~\langle \Phi _{D}\rangle _{--}$ & -0.2790 & 
-0.04360 \\ \hline
$\langle \Phi _{SS}\rangle _{++},~\langle \Phi _{SS}\rangle _{--}$ & -0.03637
& -0.00947 \\ \hline
$\langle -2\Phi _{T}\rangle _{--}$ &  & -0.04172 \\ \hline
$-6\langle \Phi _{SO}\rangle _{--}$ &  & 0.1133 \\ \hline
$2\langle \Phi _{SOT}\rangle _{--}$ &  & 0.02153 \\ \hline
$\langle \Phi _{SD}\rangle _{++},~\langle \Phi _{SD}\rangle _{--}$ & -0.3154
& 0.03461 \\ \hline\hline
$\frac{2\sqrt{2}}{3}\langle 3\Phi _{T}\rangle _{+-},\frac{2\sqrt{2}}{3}
\langle 3\Phi _{T}\rangle _{-+}~$ & -0.3088 & -0.3088 \\ \hline
$\frac{2\sqrt{2}}{3}\langle -6\Phi _{SOT}\rangle _{+-}$ & 0.6177 &  \\ 
\hline\hline
$\langle \Phi \rangle _{+-},~\langle \Phi \rangle _{-+}$ & 0.3090 & -0.3088
\\ \hline
$\langle b^{2}\rangle _{++},\langle b^{2}\rangle _{--}$ & -0.1750 & 0.3158
\\ \hline
$\langle b^{2}\rangle _{+-},\langle b^{2}\rangle _{-+}$ & 0.3090 & -0.3088
\\ \hline\hline
$\langle b^{2}\rangle _{Total}$=0.1411 &  &  \\ \hline
\end{tabular}
\caption{Expectation values of various terms in Eqs.\ (\protect\ref{rhos})
and (\protect\ref{rhod}) (in GeV$^{2}$) that contribute to the rho
eigenvalue equations (\protect\ref{rhos} and (\protect\ref{rhod})).}
\end{table}

\subsubsection{\protect\bigskip Detailed analysis of the vector meson
potentials and wave functions. \ }

As may be anticipated from the dip behavior displayed in Eq.\ (\ref{dip})
for the case of QED interactions we anticipate that a similar structure may
appear for the QCD wave functions. \ This is borne out by Figs.\ 3 and 4 for
the $S$ and $D$ state contributions to the $J/\psi$ and the $\rho$ mesons. \
In the equations below the analytic origin of the dip behavior is shown from
the behaviors of the quasipotential contributions.

\begin{figure}[ptb]
\includegraphics[angle=0,scale=0.50]{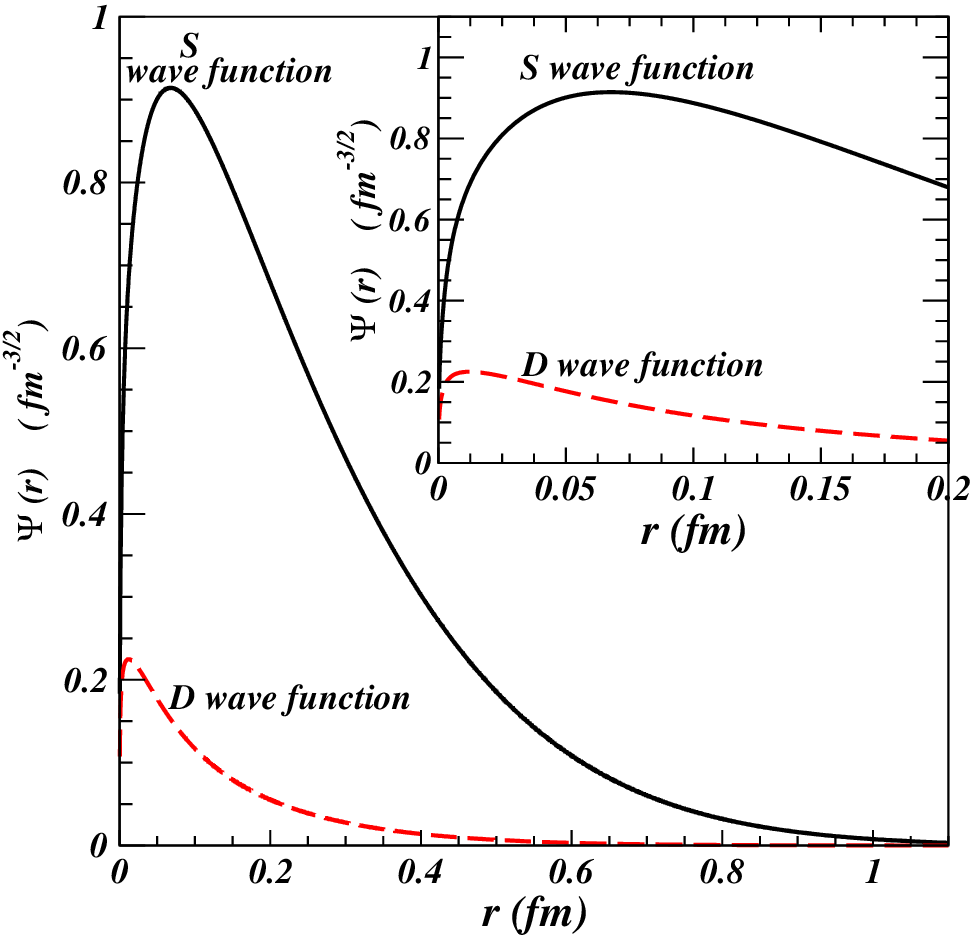} 
\caption{(Color online) Wave functions for the~$J/\protect\psi$ meson. The
insert in the upper right corner gives an expanded view of the wave
functions near the origin.}
\end{figure}

In contrast to the short distance behavior displayed in Eqs.\ (\ref{sing1})
and (\ref{sing2}) in the QED case, the short distance behavior for our
coupled QCD equations, corresponding to mesons such as the $J/\psi$ and $
\rho $ mesons is

\begin{figure}[ptb]
\includegraphics[angle=0,scale=0.50]{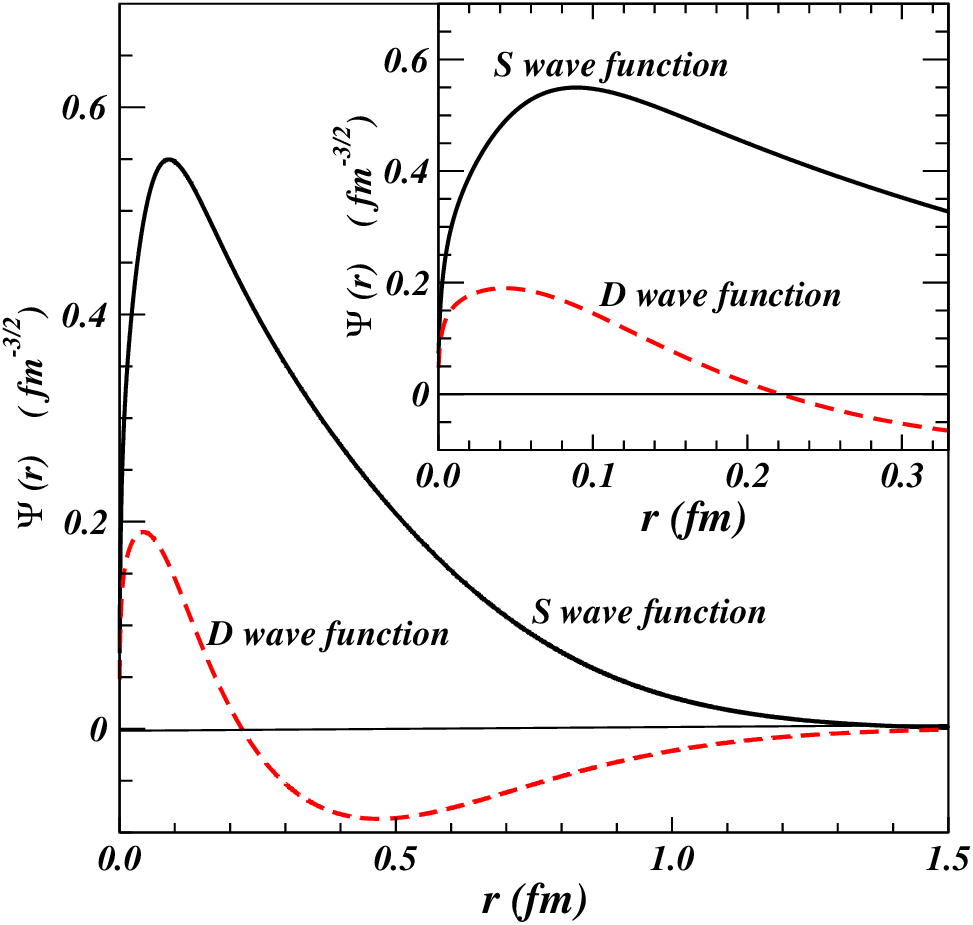} 
\caption{Wave functions for the $\protect\rho$ meson. The insert in the
upper right corner gives an expanded view of the wave functions near the
origin.}
\end{figure}

\begin{align}
& \left[ -\frac{d^{2}}{dr^{2}}-\frac{4}{3r^{2}}\sqrt{-\frac{4\pi }{27wr\ln
(\Lambda r)}}+\frac{11}{6r^{2}}\right] u_{+}+\frac{2\sqrt{2}}{3}\left[ \frac{
2}{r^{2}}\sqrt{-\frac{4\pi }{27wr\ln (\Lambda r)}}-\frac{1}{8r^{2}}\right]
u_{-}  \notag \\
& =0,
\end{align}
and 
\begin{align}
& \left[ -\frac{d^{2}}{dr^{2}}+\frac{6}{r^{2}}+\frac{4}{3r^{2}}\sqrt{-\frac{
4\pi }{27wr\ln (\Lambda r)}}-\frac{49}{12r^{2}}\right] u_{-}+\frac{2\sqrt{2}
}{3}\left[ -\frac{1}{r^{2}}\sqrt{-\frac{4\pi }{27wr\ln (\Lambda r)}}-\frac{1
}{8r^{2}}\right] u_{+}  \notag \\
& =0.
\end{align}
Let us assume a short distance behavior of 
\begin{align}
u_{+}(r)& =f(r),  \notag \\
u_{-}(r)& =\frac{f(r)}{\sqrt{2}}=\frac{u_{+}(r)}{\sqrt{2}}.
\end{align}
Then just as in the case of QED, the more singular terms (here $\sim ({1}/{
r^{2}})\sqrt{-4\pi /27wr\ln (\Lambda r)}$) cancel among themselves and we
are left with 
\begin{equation}
\left[ -\frac{d^{2}}{dr^{2}}+\frac{11}{6r^{2}}\right] f(r)+\frac{2}{3}\left[
-\frac{1}{8r^{2}}\right] f(r)=0,
\end{equation}
and 
\begin{equation}
\left[ -\frac{d^{2}}{dr^{2}}+\frac{23}{12r^{2}}\right] f(r)+\frac{4}{3}\left[
-\frac{1}{8r^{2}}\right] f(r)=0.
\end{equation}
which have the identical behavior of 
\begin{equation}
\left[ -\frac{d^{2}}{dr^{2}}+\frac{7}{4r^{2}}\right] f(r)=0,
\end{equation}
or 
\begin{equation}
f(r)\sim r^{(1/2+\sqrt{2})}
\end{equation}
The corresponding radial parts of the wave functions would be 
\begin{align}
\psi _{+}(r)& =r^{\sqrt{2}-\frac{1}{2}},  \notag \\
\psi _{-}(r)& =\frac{r^{\sqrt{2}-\frac{1}{2}}}{\sqrt{2}},  \label{asy}
\end{align}
compared to the corresponding radial parts of the QED wave functions of 
\begin{align}
\psi _{+}(r)& =r^{\sqrt{2-\alpha ^{2}}-\frac{1}{2}},  \notag \\
\psi _{-}(r)& =\frac{r^{\sqrt{2-\alpha ^{2}}-\frac{1}{2}}}{\sqrt{2}}.
\end{align}
Numerically, one finds results that approach those of Eq.\ (\ref{asy}) in
terms of the $S/D$ ratio and power behavior.

\section{Conclusion and Further Remarks}

We have shown how the TBDE of constraint dynamics handles the problem of
effective potentials that are singular in the weak potential limit. The most
noteworthy feature is that different portions of the quasipotential $\Phi $
contribute to the perturbative and nonperturbative treatment of spectral
effects. The results we found for the coupled $^{3}S_{1}$-$^{3}D_{1} $
system demonstrates that this effect extends not only to different terms of
a given equation, but also bridges the divide between the various coupled
components of the wave function and effective potential. We find
unexpectedly, that the behavior of the $J=1,$ $S$ and $D$ waves (and more
generally for an arbitrary $J$) are simply proportional very near the
origin, with a common power-law behavior.

We have introduced a new QCD potential that has many of the features of the
Adler-Piran and Richardson potentials but with a much simpler
parametrization. \ Although not giving results (rms deviation=21 MeV) as
good as the former \cite{crater2} (rms deviation=14 MeV), it nevertheless
yields a meson mass spectrum that agrees reasonably well with experiment. We
examine in particular how various contributions to the pion eigenvalue
equation can lead to a pion of a small rest mass. The detailed treatment of
the pion shows a unique feature of it relativistic behavior, namely a
behavior that superficially appears nonrelativistic but on further analysis
displays extremely large relativistic and nearly canceling contributions to
and from the potential. \ This gives some insight into how the potential
model leads to a small pion mass for small quark mass, an important
consequence of spontaneous symmetry breaking. By contrast our Table IV shows
why mesons with other quantum numbers, would not display such a small quark
mass behavior for the bound state energy. \ 

Another unexpected behavior for the QCD wave functions and effective
potentials is that at short distance, the $J=1,$ $S$ and $D$ waves for the $
J/\psi $ mesons are not significantly different from that which appears in
QED, in spite of the asymptotic freedom behavior that occurs in QCD. It may
be worthwhile to investigate what observable features of these bound states
(both in QED and QCD) may reflect these unexpected connections between the
two coupled wave functions.

In order that our primary bound relativistic Schr\"{o}dinger equation (\ref
{57} ) and the radial equations (\ref{ss})-(\ref{mi}) have an additional
ease of use for others who may wish to apply them, we have added Appendix C
to detail the relation between the $\Phi $'s that appear in (\ref{phi}) and
the invariants $A(r)$ and $S(r)$ and their derivatives.

In addition to applications to meson spectroscopy, our work has implications
for studies related to quarkonia at finite temperatures and in a quark-gluon
plasma \cite{Adc04,Ada05a,Ars04,Bac04}. Within the potential model, the
quark drip line at which quarkonia begins to be unbound has been estimated
using non-relativistic quark models \cite{Won07,Won07a}. A more definitive
investigation of the stability of quarkonium states, especially those with
light quarks, necessitates the use of the relativistic formalism developed
here. The analytic structure of the $A$ and $S$ invariants given in Eq.\ (
\ref{sa}) lend themselves to practical modeling at finite temperatures,
while giving an adequate account of the zero temperature limit. The
relativistic one developed here provides a more realistic model than the
most often used Cornell-type linear-plus color-Coulomb for the case of $T=0.$
It will be of interest to investigate the composite properties of the plasma
and to determine the region of temperatures in which different quarkonia
become unbound using a temperature dependent extension of the simple quark
model presented in this paper. An important problem for future work in
relation to the present paper is to determine how best to apportion these
temperature dependent potentials between the $A(r)$ and $S(r)$ invariants.

This research was supported in part by the Division of Nuclear Physics, U.S.
Department of Energy, under Contract No. DE-AC05-00OR22725, managed by
UT-Battelle, LLC and by the Korea Science and Engineering Foundation (KOSEF)
grant funded by the Korean government (MEST) (2006-8-0083).

\bigskip 
\begin{appendix}

\section{Relativistic Schr\"{o}dinger Equation Details}

Here we present an outline of some details of Eq. (\ref{57}) given in full
elsewhere (see \cite{cra87,bckr,cra94,crater2,cra82,saz94} and works cited
therein and \cite{unusual}). \ For classical \cite{fw} or quantum field
theories \cite{saz97} for separate scalar and vector interactions one can
show that the spin independent part of the quasipotential $\Phi _{w}~$
involves the difference of squares of the invariant mass and energy
potentials ($M_{i}$ and $E_{i}$ respectively) \ 
\begin{align}
M_{i}^{2}& =m_{i}^{2}+2m_{w}S+S^{2};\ E_{i}^{2}=\varepsilon
_{i}^{2}-2\varepsilon _{w}A+A^{2},  \label{kg1} \\
M_{i}^{2}-E_{i}^{2}& =2m_{w}S+S^{2}+2\varepsilon _{w}A-A^{2}-b^{2}(w).
\label{kg}
\end{align}
where 
\begin{equation}
m_{w}=\frac{m_{1}m_{2}}{w}\ \ ,~~\varepsilon _{w}=\frac{
(w^{2}-m_{1}^{2}-m_{2}^{2})}{2w},  \label{emw}
\end{equation}
are respectively the relativistic reduced mass and energy of the fictitious
particle of relative motion introduced by Todorov \cite{tod71,tod76} and
satisfy the effective one-body Einstein condition 
\begin{equation}
\epsilon _{w}^{2}-m_{w}^{2}=b^{2}(w).
\end{equation}

Eqs.\ (\ref{tbdea}) and (\ref{tbdeb}) contain an important hidden hyperbolic
structure \cite{jmp}. To reveal and employ it one introduces two independent
invariant functions $L(x_{\perp })$ and $\mathcal{G}(x_{\perp })$, in terms
of which the invariant mass and energy potentials take the forms: 
\begin{align}
M_{1}& =m_{1}\ \cosh L(S,A)\ +m_{2}\sinh L(S,A),  \notag \\
M_{2}& =m_{2}\ \cosh L(S,A)\ +m_{1}\ \sinh L(S,A),  \label{scl}
\end{align}
\begin{align}
E_{1}& =\varepsilon _{1}\ \cosh \mathcal{G}(A)\ -\varepsilon _{2}\sinh 
\mathcal{G}(A),  \notag \\
E_{2}& =\varepsilon _{2}\ \cosh \mathcal{G}(A)-\varepsilon _{1}\ \sinh 
\mathcal{G}(A).  \label{vectl}
\end{align}
Strictly speaking, the forms in Eq.\ (\ref{kg1}) are for scalar and
time-like vector interactions. Eq. (\ref{57}) below involves combined scalar
and electromagnetic-like vector interactions (this amounts to working in the
Feynman gauge with the simplest relation between space- and time-like parts,
see \cite{cra88,crater2})). \ In that case the\ mass and energy potentials in
place of Eq.\ (\ref{kg1}) are respectively 
\begin{equation}
M_{i}^{2}=m_{i}^{2}+\exp (2\mathcal{G)(}2m_{w}S\mathcal{+}S^{2}),
\label{one}
\end{equation}
so that 
\begin{equation}
\exp (L(S,A))=\frac{\sqrt{m_{1}^{2}+\exp (2\mathcal{G)(}2m_{w}S\mathcal{+}
S^{2})}+\sqrt{m_{2}^{2}+\exp (2\mathcal{G)(}2m_{w}S\mathcal{+}S^{2})}}{
m_{1}+m_{2}},\   \label{1.5}
\end{equation}
and 
\begin{equation}
E_{i}^{2}=\exp (2\mathcal{G)(\varepsilon }_{i}-A)^{2},  \label{two}
\end{equation}
with$~$ 
\begin{equation}
\exp (2\mathcal{G(}A\mathcal{))=}\frac{1}{(1-2A/w)}\equiv G^{2}.
\label{three}
\end{equation}
In terms of $\mathcal{G}$ and the constituent momenta $p_{1}$ and $p_{2}$,
the individual four-vector potentials take the forms\cite{crater2} 
\begin{align}
A_{1}& =[1-\mathrm{\cosh }(\mathcal{G})]p_{1}+\mathrm{\sinh }(\mathcal{G}
)p_{2}-\frac{i}{2}(\partial \exp \mathcal{G}\cdot \gamma _{2})\gamma _{2}, 
\notag \\
A_{2}& =[1-\mathrm{\cosh }(\mathcal{G})]p_{2}+\mathrm{\sinh }(\mathcal{G}
)p_{1}+\frac{i}{2}(\partial \exp \mathcal{G}\cdot \gamma _{1})\gamma _{1}.
\label{vec}
\end{align}
In terms of the three sets of invariants (\ref{one})-(\ref{three}) the
coupled TBDE (\ref{tbdea}) and (\ref{tbdeb}) then take the form 
\begin{align}
S_{1}\psi & =\big(-G\beta _{1}\Sigma _{1}\cdot \mathcal{P}_{2}+E_{1}\beta
_{1}\gamma _{51}+M_{1}\gamma _{51}-G\frac{i}{2}\Sigma _{2}\cdot \partial (
\mathcal{G}\beta _{1}+L\beta _{2})\gamma _{51}\gamma _{52}\big)\psi =0, 
\notag \\
\mathcal{S}_{2}\psi & =\big(G\beta _{2}\Sigma _{2}\cdot \mathcal{P}
_{1}+E_{2}\beta _{2}\gamma _{52}+M_{2}\gamma _{52}+G\frac{i}{2}\Sigma
_{1}\cdot \partial (\mathcal{G}\beta _{2}+L\beta _{1})\gamma _{51}\gamma
_{52}\big)\psi =0,  \label{b2}
\end{align}
in which 
\begin{equation}
\mathcal{P}_{i}\equiv p-\frac{i}{2}\Sigma _{i}\cdot \partial \mathcal{G}
\Sigma _{i},
\end{equation}
depending on gamma matrices with standard block forms (see Eq.\ (2.28) in 
\cite{crater2} for their explicit forms) and where 
\begin{equation}
\Sigma _{i}=\gamma _{5i}\beta _{i}\gamma _{\perp i}.
\end{equation}
The Klein-Gordon like potential energy terms appearing in the Pauli form (
\ref{57}) of Eq.\ (\ref{b2}) are as in Eq.\ (\ref{kg}).

To obtain the simple Pauli form of Eq.\ (\ref{paul1}) involves steps
analogous to those used in Eqs.\ (\ref{7})-(\ref{9}) but with the
combinations $\phi _{\pm }=\psi _{1}\pm \psi _{4}$ and $\chi _{\pm }=\psi
_{2}\pm \psi _{3}$ instead of the the individual $\psi _{i}$. This allows
the Pauli forms to reduce to 4 uncoupled 4 component relativistic
Schrodinger equations \cite{saz94,cra94,long,crater2,liu}. \ Working in the
c.m. frame in which $\hat{P}=(1,\mathbf{0)}$ and $\hat{r}=(0,\mathbf{\hat{r})
}$ and then further defining four component wave functions $\psi _{\pm
},\eta _{\pm }$ related to the above by \cite{liu} 
\begin{align}
\phi _{\pm }& =\exp (F+K\boldsymbol{\sigma}_{1}\mathbf{\cdot \hat{r}}
\boldsymbol{\sigma}_{2}\mathbf{\cdot \hat{r}})\psi _{\pm }=(\exp F)(\cosh
K+\sinh K\boldsymbol{\sigma}_{1}\mathbf{\cdot \hat{r}}\boldsymbol{\sigma}_{2}
\mathbf{\cdot \hat{r}})\psi _{\pm },  \notag \\
\chi _{\pm }& =\exp (F+K\boldsymbol{\sigma}_{1}\mathbf{\cdot \hat{r}}
\boldsymbol{\sigma}_{2}\mathbf{\cdot \hat{r}})\eta _{\pm }=(\exp F)(\cosh
K+\sinh K\boldsymbol{\sigma}_{1}\mathbf{\cdot \hat{r}}\boldsymbol{\sigma}_{2}
\mathbf{\cdot \hat{r}})\eta _{\pm }.  \label{fk}
\end{align}
in which 
\begin{align}
F& =\frac{1}{2}\log \frac{\mathcal{D}}{\varepsilon _{2}m_{1}+\varepsilon
_{1}m_{2}}-\mathcal{G},  \notag \\
\mathcal{D}& \mathcal{=}E_{2}M_{1}+E_{1}M_{2},  \notag \\
K& =\frac{(\mathcal{G}+L)}{2},  \label{kf}
\end{align}
will yield equations which in the limit when one mass becomes extremely
large reduce to the Schr\"{o}dinger or Pauli-forms discussed in 
Section III. In analogy to what occurred there the decoupled form of the Schr 
\"{o}dinger equation for $\psi _{+}$ has the convenient property that \ the
coefficients of the first order relative momentum terms vanish.

\ Using the results in \cite{liu} we obtain for the general case of unequal
masses the relativistic Schr\"{o}dinger equation (\ref{57}) that is a
detailed c.m.\ form of Eq.\ (\ref{paul1}). In that equations we have
introduced the abbreviations 
\begin{align}
\Phi _{D}& =-\frac{2F^{\prime }(\cosh 2K-1)}{r}+F^{\prime 2}+K^{\prime 2}+ 
\frac{2K^{\prime }\sinh 2K}{r}-\boldsymbol{\nabla}^{2}F  \notag \\
& -\frac{2(\cosh 2K-1)}{r^{2}}+m(r),  \notag \\
\Phi _{SO}& =-\frac{F^{\prime }}{r}-\frac{F^{\prime }(\cosh 2K-1)}{r}-\frac{
(\cosh 2K-1)}{r^{2}}+\frac{K^{\prime }\sinh 2K}{r},  \notag \\
\Phi _{SOD}& =(l^{\prime }\cosh 2K-q^{\prime }\sinh 2K),  \notag \\
\Phi _{SOX}& =(q^{\prime }\cosh 2K-l^{\prime }\sinh 2K),  \notag \\
\Phi _{SS}& =k(r)+\frac{2K^{\prime }\sinh 2K}{3r}-\frac{2F^{\prime }(\cosh
2K-1)}{3r}-\frac{2(\cosh 2K-1)}{3r^{2}}  \notag \\
& +\frac{2F^{\prime }K^{\prime }}{3}-\frac{\boldsymbol{\nabla}^{2}K}{3}, \\
\Phi _{T}& =\frac{1}{3}[n(r)+\frac{3F^{\prime }\sinh 2K}{r}+\frac{F^{\prime
}(\cosh 2K-1)}{r}+2F^{\prime }K^{\prime }-\frac{K^{\prime }\sinh 2K}{r}- 
\frac{3K^{\prime }(\cosh 2K-1)}{r}  \notag \\
& -\boldsymbol{\nabla}^{2}K+\frac{3\sinh 2K}{r^{2}}+\frac{(\cosh 2K-1)}{
r^{2} }],  \notag \\
\Phi _{SOT}& =-K^{\prime }\frac{\cosh 2K-1}{r}+\frac{\sinh 2K}{r^{2}}-\frac{
K^{\prime }}{r}+\frac{F^{\prime }\sinh 2K}{r}.  \label{phi}
\end{align}
in which 
\begin{align}
k(r)& =\frac{1}{3}\nabla ^{2}(K+\mathcal{G)}-\frac{2F^{\prime }(\mathcal{G}
^{\prime }+K^{\prime })}{3}\mathcal{-}\frac{1}{2}\mathcal{G}^{\prime 2} 
\notag \\
n(r)& =\frac{1}{3}[\nabla ^{2}K-\frac{1}{2}\nabla ^{2}\mathcal{G}+\frac{3( 
\mathcal{G}^{\prime }-2K^{\prime })}{2r}+F^{\prime }(\mathcal{G}^{\prime
}-2K^{\prime })],  \notag \\
m(r)& =-\frac{1}{2}\nabla ^{2}\mathcal{G+}\frac{3}{4}\mathcal{G}^{\prime 2}+ 
\mathcal{G}^{\prime }F^{\prime }-K^{\prime 2},  \label{knm}
\end{align}
and 
\begin{align}
l^{\prime }(r)& =-\frac{1}{2r}\frac{E_{2}M_{2}-E_{1}M_{1}}{
E_{2}M_{1}+E_{1}M_{2}}(L-\mathcal{G})^{\prime },  \notag \\
q^{\prime }(r)& =\frac{1}{2r}\frac{E_{1}M_{2}-E_{2}M_{1}}{
E_{2}M_{1}+E_{1}M_{2}}(L-\mathcal{G})^{\prime }.
\end{align}
(The prime symbol stands for $d/dr)$. For $L=J$ states, the hyperbolic
terms cancel and the spin-orbit difference terms in general produce
spin mixing except for equal masses or $J=0$. For utility of use we
have listed in Appendix C the explicit forms that appear in the above
$\Phi $'s in terms of the general invariant potentials $A(r)$ and
$S(r).~$\ The radial components of Eq. (\ref{57}) are given in
Appendix B.

\section{Radial Equations}

The following are radial eigenvalue equations corresponding to Eq.\ (\ref{57}
) \cite{liu}. For a general singlet $^{1}J_{J}$ wave function $v_{0}$
coupled to a general triplet $^{3}J_{J}$ wave function $v_{1}$, the wave
equation

\begin{align}
& \{-\frac{d^{2}}{dr^{2}}+\frac{J(J+1)}{r^{2}}+2m_{w}S+S^{2}+2\varepsilon
_{w}A-A^{2}+\Phi _{D}\mathbf{-}3\Phi _{SS}\}v_{0}  \notag \\
& +2\sqrt{J(J+1)}(\Phi _{SOD}-\Phi _{SOX})v_{1}  \notag \\
& =b^{2}v_{0},  \label{ss}
\end{align}
is coupled to

\begin{align}
& \{-\frac{d^{2}}{dr^{2}}+\frac{J(J+1)}{r^{2}}+2m_{w}S+S^{2}+2\varepsilon
_{w}A-A^{2}+\Phi _{D}  \notag \\
& -2\Phi _{SO}+\Phi _{SS}+2\Phi _{T}-2\Phi _{SOT}\}v_{1}+2\sqrt{J(J+1)}(\Phi
_{SOD}+\Phi _{SOX})v_{0}  \notag \\
& =b^{2}v_{1}.  \label{pp}
\end{align}
For a general $S=1,$ $J=L+1$ \ wave function $u_{+}~$coupled to general $
S=1, $ $J=L-1~$wave function $u_{-}$ \ the equation

\begin{align}
& \{-\frac{d^{2}}{dr^{2}}+\frac{J(J-1)}{r^{2}}+2m_{w}S+S^{2}+2\varepsilon
_{w}A-A^{2}+\Phi _{D}  \notag \\
& +2(J-1)\Phi _{SO}+\Phi _{SS}+\frac{2(J-1)}{2J+1}(\Phi _{SOT}-\Phi
_{T})\}u_{+}  \notag \\
& +\frac{2\sqrt{J(J+1)}}{2J+1}\{3\Phi _{T}-2(J+2)\Phi _{SOT}\}u_{-}  \notag
\\
& =b^{2}u_{+},  \label{pl}
\end{align}
$\allowbreak \allowbreak \allowbreak $is coupled to 
\begin{align}
& \{-\frac{d^{2}}{dr^{2}}+\frac{(J+1)(J+2)}{r^{2}}+2m_{w}S+S^{2}+2
\varepsilon _{w}A-A^{2}+\Phi _{D}  \notag \\
& -2(J+2)\Phi _{SO}+\Phi _{SS}+\frac{2(J+2)}{2J+1}(\Phi _{SOT}-\Phi
_{T})\}u_{-}  \notag \\
& +\frac{2\sqrt{J(J+1)}}{2J+1}\{3\Phi _{T}+2(J-1)\Phi _{SOT}\}u_{+}  \notag
\\
& =b^{2}u_{-}.  \label{mi}
\end{align}

\bigskip

\section{Explicit Expressions for terms in the Relativistic Schr\"{o}dinger
Equation (\protect\ref{57}) from $A(r)$ and $S(r)$}

Given the functions $A(r)$ and $S(r)$ for the interaction, users of the
relativistic Schr\"{o}dinger equation (\ref{57}) will find it convenient to
have an explicit expression in an order that would be useful for programing
the terms in the associated equation (\ref{phi}). We use the definitions
given in Eqs.\ (\ref{kf}), (\ref{knm}), and (\ref{one} )-( \ref{three}). In
order that the terms in Eq.\ (\ref{phi}) be reduced to expressions involving
just $A(r),~S(r)$ and their derivatives, we list the following formulae: 
\begin{eqnarray}
F^{\prime } &=&\frac{(L^{\prime }-\mathcal{G}^{\prime
})(E_{2}M_{2}+E_{1}M_{1})}{2(E_{2}M_{1}+E_{1}M_{2})}-\mathcal{G}^{\prime }, 
\notag \\
E_{1} &=&\frac{\varepsilon _{1}-A}{\sqrt{(w-2A)/w}},~E_{2}=\frac{\varepsilon
_{2}-A}{\sqrt{(w-2A)/w}},  \notag \\
M_{1} &=&\sqrt{m_{1}^{2}+\frac{2m_{w}S+S^{2}}{(w-2A)/w}},~M_{2}=\sqrt{
m_{2}^{2}+\frac{2m_{w}S+S^{2}}{(w-2A)/w}}~,  \notag \\
L^{\prime } &=&\frac{M_{1}^{\prime }}{M_{2}}=\frac{M_{2}^{\prime }}{M_{1}}= 
\frac{w}{M_{1}M_{2}}\left( \frac{S^{\prime }(m_{w}+S)}{w-2A}+\frac{
(2m_{w}S+S^{2})A^{\prime }}{(w-2A)^{2}}\right) ,  \notag \\
\mathcal{G}^{\prime } &=&\frac{A^{\prime }}{w-2A}.
\end{eqnarray}
Also needed are 
\begin{eqnarray}
\cosh 2K &=&\frac{1}{2}\left( \frac{(\varepsilon _{1}+\varepsilon
_{2})(M_{1}+M_{2})}{(m_{1}+m_{2})(E_{1}+E_{2})}+\frac{
(m_{1}+m_{2})(E_{1}+E_{2})}{(\varepsilon _{1}+\varepsilon _{2})(M_{1}+M_{2})}
\right) ,  \notag \\
\sinh 2K &=&\frac{1}{2}\left( \frac{(\varepsilon _{1}+\varepsilon
_{2})(M_{1}+M_{2})}{(m_{1}+m_{2})(E_{1}+E_{2})}-\frac{
(m_{1}+m_{2})(E_{1}+E_{2})}{(\varepsilon _{1}+\varepsilon _{2})(M_{1}+M_{2})}
\right) ,
\end{eqnarray}
and 
\begin{eqnarray}
K^{\prime } &=&\frac{\mathcal{G}^{\prime }+L^{\prime }}{2},  \notag \\
\boldsymbol{\nabla}^{2}F &=&\frac{(\boldsymbol{\nabla}^{2}L- 
\boldsymbol{\nabla}^{2}\mathcal{G})(E_{2}M_{2}+E_{1}M_{1})}{
2(E_{2}M_{1}+E_{1}M_{2})}-(L^{\prime }-\mathcal{G}^{\prime })^{2}\frac{
(m_{1}^{2}-m_{2}^{2})^{2}}{2\left( E_{2}M_{1}+E_{1}M_{2}\right) ^{2}}- 
\boldsymbol{\nabla}^{2}\mathcal{G},  \notag \\
\boldsymbol{\nabla}^{2}L &=&\frac{-L^{\prime 2}(M_{1}^{2}+M_{2}^{2})}{
M_{1}M_{2}}  \notag \\
&&+\frac{w}{M_{1}M_{2}}\left( \frac{\boldsymbol{\nabla}^{2}S(m_{w}+S)+S^{
\prime 2}}{w-2A}+\frac{4S^{\prime }(m_{w}+S)A^{\prime }+(2m_{w}S+S^{2}) 
\boldsymbol{\nabla}^{2}A}{(w-2A)^{2}}+\frac{4(2m_{w}S+S^{2})A^{\prime 2}}{
(w-2A)^{3}}\right) ,  \notag \\
\boldsymbol{\nabla}^{2}\mathcal{G} &=&\frac{\boldsymbol{\nabla}^{2}A}{w-2A}
+2 \mathcal{G}^{\prime 2}.
\end{eqnarray}
The expressions for $k(r),m(r),$ and $n(r)$ that appear in Eq.\ (\ref{phi})
are given in Eq.\ (\ref{knm}). They can be evaluated using the above
expressions plus 
\begin{equation}
\boldsymbol{\nabla}^{2}K=\frac{\boldsymbol{\nabla}^{2}\mathcal{G+} 
\boldsymbol{\nabla}^{2}L}{2}.
\end{equation}
The only remaining parts of Eq.\ (\ref{phi}) that need expressing are those
for $l^{\prime }$ and $q^{\prime }.$ Using Eq.\ (\ref{kf}) they can be
obtained in terms of the above formulae.\vspace*{0.3cm}

\end{appendix}


\begin{thebibliography}{99}
\bibitem{Err08} There are inadvertent typographical errors in Eqs. (3.21)
and (3.22) in Ref.\ \cite{decay}. In the fourth line of Eq. (3.21) inside
the square bracket for $\mathbf{\sigma }_{1}\cdot \mathbf{\sigma }_{2}$, the
term $-{F^{\prime }\sinh 2K}/{2r}$ should read $-{F^{\prime }\sinh 2K}/{r}$
and the term $+{\sinh 2K}/{r^{2}}$ on the next line should read $-{\sinh 2K}/
{r^{2}}$. In the first equation of Eq.\ (3.22), the term $-\mathcal{G}
^{\prime }K^{\prime }/2$ should read $-\mathcal{G}^{\prime }F^{\prime }$,
and in the second equation of Eq.\ (3.22), the term $-3\mathcal{G}^{\prime
}/2r$ should read $+3\mathcal{G}^{\prime }/2r-3K^{\prime }/r$.
\end{thebibliography}
\end{document}